\documentclass{emulateapj}
\usepackage{apjfonts}

\newcommand{\etal}{et~al.}
\newcommand{\PVdblt}{{\rm P}\kern 0.1em{\sc v}~$\lambda\lambda 1117, 1128$}
\newcommand{\CaIIdblt}{{\rm Ca}\kern 0.1em{\sc ii}~$\lambda\lambda 3934, 3969$}
\newcommand{\AlIIIdblt}{{\rm Al}\kern 0.1em{\sc iv}~$\lambda\lambda 1855, 1863$}
\newcommand{\CIVdblt}{{\rm C}\kern 0.1em{\sc iv}~$\lambda\lambda 1548, 1550$}
\newcommand{\MgIIdblt}{{\rm Mg}\kern 0.1em{\sc ii}~$\lambda\lambda 2796, 2803$}
\newcommand{\NVdblt}{{\rm N}\kern 0.1em{\sc v}~$\lambda\lambda 1238, 1242$}  
\newcommand{\SVIdblt}{{\rm S}\kern 0.1em{\sc vi}~$\lambda\lambda 933, 944$} 
\newcommand{\OVIdblt}{{\rm O}\kern 0.1em{\sc vi}~$\lambda\lambda 1031, 1037$} 
\newcommand{\SiIIdblt}{{\rm Si}\kern 0.1em{\sc ii}~$\lambda\lambda 1190, 1193$} 
\newcommand{\SiIVdblt}{{\rm Si}\kern 0.1em{\sc iv}~$\lambda\lambda 1393, 1402$} 
\newcommand{\PV}{\hbox{{\rm P}\kern 0.1em{\sc v}}}
\newcommand{\AlI}{\hbox{{\rm Al}\kern 0.1em{\sc i}}}
\newcommand{\AlII}{\hbox{{\rm Al}\kern 0.1em{\sc ii}}}
\newcommand{\AlIII}{{\hbox{\rm Al}\kern 0.1em{\sc iii}}}
\newcommand{\CaII}{\hbox{{\rm Ca}\kern 0.1em{\sc ii}}}
\newcommand{\CII}{\hbox{{\rm C}\kern 0.1em{\sc ii}}}
\newcommand{\CIIe}{\hbox{{\rm C$^{\ast}$}\kern 0.1em{\sc ii}}}
\newcommand{\CIII}{\hbox{{\rm C}\kern 0.1em{\sc iii}}}
\newcommand{\CIV}{\hbox{{\rm C}\kern 0.1em{\sc iv}}}
\newcommand{\CV}{\hbox{{\rm C}\kern 0.1em{\sc v}}}
\newcommand{\HI}{\hbox{{\rm H}\kern 0.1em{\sc i}}}
\newcommand{\HII}{\hbox{{\rm H}\kern 0.1em{\sc ii}}}
\newcommand{\Lya}{\hbox{{\rm Ly}\kern 0.1em$\alpha$}}
\newcommand{\Lyb}{\hbox{{\rm Ly}\kern 0.1em$\beta$}}
\newcommand{\Lyg}{\hbox{{\rm Ly}\kern 0.1em$\gamma$}}
\newcommand{\Lyd}{\hbox{{\rm Ly}\kern 0.1em$\delta$}}
\newcommand{\Lye}{\hbox{{\rm Ly}\kern 0.1em$\epsilon$}}
\newcommand{\Lyphi}{\hbox{{\rm Ly}\kern 0.1em$\phi$}}
\newcommand{\Lyfive}{\hbox{{\rm Ly}\kern 0.1em$5$}}
\newcommand{\Lysix}{\hbox{{\rm Ly}\kern 0.1em$6$}}
\newcommand{\Lyseven}{\hbox{{\rm Ly}\kern 0.1em$7$}}
\newcommand{\Lyeight}{\hbox{{\rm Ly}\kern 0.1em$8$}}
\newcommand{\Lynine}{\hbox{{\rm Ly}\kern 0.1em$9$}}
\newcommand{\Lyten}{\hbox{{\rm Ly}\kern 0.1em$10$}}
\newcommand{\Lyeleven}{\hbox{{\rm Ly}\kern 0.1em$11$}}
\newcommand{\HeI}{\hbox{{\rm He}\kern 0.1em{\sc i}}}
\newcommand{\HeII}{\hbox{{\rm He}\kern 0.1em{\sc ii}}}
\newcommand{\FeI}{\hbox{{\rm Fe}\kern 0.1em{\sc i}}}
\newcommand{\FeII}{\hbox{{\rm Fe}\kern 0.1em{\sc ii}}}
\newcommand{\FeIII}{\hbox{{\rm Fe}\kern 0.1em{\sc iii}}}
\newcommand{\MnII}{\hbox{{\rm Mn}\kern 0.1em{\sc ii}}}
\newcommand{\MgI}{\hbox{{\rm Mg}\kern 0.1em{\sc i}}}
\newcommand{\MgII}{\hbox{{\rm Mg}\kern 0.1em{\sc ii}}}
\newcommand{\MgIII}{\hbox{{\rm Mg}\kern 0.1em{\sc iii}}}
\newcommand{\NI}{\hbox{{\rm N}\kern 0.1em{\sc i}}}
\newcommand{\NII}{\hbox{{\rm N}\kern 0.1em{\sc ii}}}
\newcommand{\NIII}{\hbox{{\rm N}\kern 0.1em{\sc iii}}}
\newcommand{\NV}{\hbox{{\rm N}\kern 0.1em{\sc v}}}
\newcommand{\OVI}{\hbox{{\rm O}\kern 0.1em{\sc vi}}}
\newcommand{\OI}{\hbox{{\rm O}\kern 0.1em{\sc i}}}
\newcommand{\OII}{\hbox{[{\rm O}\kern 0.1em{\sc ii}]}}
\newcommand{\OIII}{\hbox{[{\rm O}\kern 0.1em{\sc iii}]}}
\newcommand{\OIV}{\hbox{{\rm O}\kern 0.1em{\sc iv}]}}
\newcommand{\SI}{{\rm S}\kern 0.1em{\sc i}}
\newcommand{\SIV}{{\rm S}\kern 0.1em{\sc iv}}
\newcommand{\SVI}{{\rm S}\kern 0.1em{\sc vi}}
\newcommand{\SiI}{\hbox{{\rm Si}\kern 0.1em{\sc i}}}
\newcommand{\SiII}{\hbox{{\rm Si}\kern 0.1em{\sc ii}}}
\newcommand{\SiIII}{\hbox{{\rm Si}\kern 0.1em{\sc iii}}}
\newcommand{\SiIV}{\hbox{{\rm Si}\kern 0.1em{\sc iv}}}
\newcommand{\SII}{\hbox{{\rm S}\kern 0.1em{\sc ii}}}
\newcommand{\SIII}{\hbox{{\rm S}\kern 0.1em{\sc iii}}}
\newcommand{\NaI}{\hbox{{\rm Na}\kern 0.1em{\sc i}}}
\newcommand{\TiII}{\hbox{{\rm Ti}\kern 0.1em{\sc ii}}}
\newcommand{\kms}{\hbox{km~s$^{-1}$}}
\newcommand{\cmsq}{\hbox{cm$^{-2}$}}

\slugcomment{Accepted December 10th, 2009}

\shorttitle{\sc Galaxy Disk and Halo Gas Kinematics: Observations and Simulations}
\shortauthors{\sc Kacprzak {\etal}}

\begin{document}


\title{Halo Gas and Galaxy Disk Kinematics Derived from Observations and
$\Lambda$CDM Simulations of {\MgII} Absorption Selected Galaxies at
Intermediate Redshift}


\author{\sc
Glenn G. Kacprzak\altaffilmark{1,2},
Christopher W. Churchill\altaffilmark{2},
Daniel Ceverino\altaffilmark{3,2}, \\
Charles C. Steidel\altaffilmark{4},
Anatoly Klypin\altaffilmark{2},
and
Michael T. Murphy\altaffilmark{1}
}
                                                                                
\altaffiltext{1}{Swinburne University of Technology, Victoria 3122,
Australia {\tt gkacprzak@astro.swin.edu.au, mmurphy@astro.swin.edu.au}}

\altaffiltext{2}{New Mexico State University, Las Cruces, NM 88003
{\tt cwc@nmsu.edu, aklypin@nmsu.edu}}

\altaffiltext{3}{The Hebrew University, Jerusalem 91904, Israel 
{\tt ceverino@phys.huji.ac.il}}
 
\altaffiltext{4}{Caltech, Pasadena, CA 91125
{\tt ccs@astro.caltech.edu}}

\begin{abstract}

We obtained ESI/Keck rotation curves of 10 {\MgII} absorption selected
galaxies ($0.3 \leq z \leq 1.0$) for which we have WFPC--2/{\it HST\/}
images and high resolution HIRES/Keck and UVES/VLT quasar spectra of
the {\MgII} absorption profiles.  We perform a kinematic comparison of
these galaxies and their associated halo {\MgII} absorption.  For all
10 galaxies, the majority of the absorption velocities lie in the
range of the observed galaxy rotation velocities.  In 7/10 cases, the
absorption velocities reside fully to one side of the galaxy systemic
velocity and usually align with one arm of the rotation curve.  In all
cases, a constant rotating thick--disk model poorly reproduces the
{\it full} spread of observed {\MgII} absorption velocities when
reasonably realistic parameters are employed.  In 2/10 cases, the
galaxy kinematics, star formation surface densities, and absorption
kinematics have a resemblance to those of high redshift galaxies
showing strong outflows. We find that {\MgII} absorption velocity
spread and optical depth distribution may be dependent on galaxy
inclination. To further aid in the spatial--kinematic relationships of
the data, we apply quasar absorption line techniques to a galaxy
($v_c=180$~km/s) embedded in $\Lambda$CDM simulations.  In the
simulations, {\MgII} absorption selects metal enriched ``halo'' gas
out to $\sim100$~kpc from the galaxy, tidal streams, filaments, and
small satellite galaxies.  Within the limitations inherent in the
simulations, the majority of the simulated {\MgII} absorption arises
in the filaments and tidal streams and is infalling towards the galaxy
with velocities between $-200 \leq v_r \leq -180$~{\kms}.  The {\MgII}
absorption velocity offset distribution (relative to the simulated
galaxy) spans $\sim200$~{\kms} with the lowest frequency of detecting
{\MgII} at the galaxy systematic velocity.

\end{abstract}




\section{Introduction}

In a cosmological context, galaxy formation occurs via accretion of
gas from the cosmic web and from galaxy--galaxy mergers. The galaxy
kinematics reflect these processes.  As galaxies evolve, a complex
interplay develops between the star formation, which can generate
winds, and an array of kinematic structures such as tidal streams,
galactic fountains, and filamentary infall that comprise an extended
gaseous halo.  This overall picture is suggested by observations and
$\Lambda$CDM cosmological simulations of galaxy formation. However, we
lack a thorough understanding, both observationally and theoretically,
of how these processes precisely affect the dynamics of galaxies and
their extended halos. High quality, detailed observations are required
to further develop this working scenario and produce a comprehensive
model of galaxy evolution in the cosmological context.

Observations of local galaxies provide detailed views of gas disks and
the inner $\sim25$~kpc of their halos.  \citet{oosterloo07} obtained
deep {\HI} observations of NGC~891 and found lagging, differentially
rotating halo gas kinematics with velocities decreasing with distance
above the galaxy plane.  Lagging halos are observed in several other
local galaxies \citep[e.g.,][]{sancisi01, swaters97,rand00,heald07a}.
There are also cases where halo gas has been detected with velocities
opposite to galaxy rotation \citep{fraternali01,oosterloo07}.

Evidence for the accretion of cold gas, which may lower angular
momentum and play a role in the development of lagging halos, has also
been rapidly accumulating.  Several galaxies are observed to have
galactic fountains, and be surrounded by {\HI} cloud complexes, minor
merger tidal tails, and IGM filaments
\citep[e.g.,][]{heald07a,sancisi08}.

Quasar absorption lines provide powerful probes of halo gas kinematics
to large galactocentric distances. In three galaxies, \citet{cote05}
found that low column density {\Lya} absorption does not follow galaxy
rotation to $D=390$~kpc. They suggested that the gas arises from the
cosmic web. \citet{bowen02} found that {\Lya} absorption strength
correlates with galaxy over density. At large distances, the halo gas
kinematics may no longer be coupled to the galaxy kinematics, but
reflect the motions of cosmic web.

Quasar absorption line studies of the {\MgII}
$\lambda\lambda2796,2803$ doublet produced by gaseous halos of
foreground galaxies
\citep[e.g.,][]{bb91,lebrun93,sdp94,csv96,cwc-china,zibetti06,kacprzak07}
probe the low ionization metal enriched gas--galaxy dynamics and IGM
interface.  With {\MgII} absorption lines, we can study the kinematic
conditions of galactic halos over a wide range of redshifts out to
projected galactocentric radii of several hundred kpc. {\MgII}
absorption arises in low ionization, metal enriched gas with neutral
hydrogen column densities of $10^{16} \lesssim \hbox{N(\HI)} \lesssim
10^{22}$~{\cmsq}, and thus selects a large dynamic range of gas
structures in the environments associated with galaxies
\citep{weakII,archiveI}.

The idea that {\MgII} absorbers could arise from spherical infall,
disk--like rotation, or both, is also a topic of much interest. In a
small sample of high resolution {\MgII} absorption profiles,
\citet{lanzetta92} inferred that rotation kinematics dominated at
smaller impact parameters, whereas infall kinematics dominated with
increasing impact parameter.  Armed with a larger high resolution
sample, \citet{charlton98} applied statistical tests to a variety of
kinematic models and concluded that pure disk rotation and pure halo
infall models are ruled out. However, models with contributions from
both disk rotation and spherical infall statistically reproduced
absorption profiles consistent with observed kinematics.

Among the more extreme structures, are galactic winds generated from
star forming galaxies.  At $z\leq1.4$, \citet{tremonti07} detected
{\MgII} blueshifted 500--2000~{\kms} relative to post-starburst host
galaxies and \citet{weiner08} found $300-1000$~{\kms} blueshifts in
$z\sim1.4$ star forming galaxies.  In other ionic species, similar
outflows have been observed at $z \sim 3$
\citep{pettini01,shapley03,steidel03,simcoe06,cabanac08}.  None of
these surveys have studied the dynamics of the galaxies themselves,
which are clearly needed in order to obtain a full picture of the
galaxy--halo dynamics.

A direct comparison of the galaxy disk kinematics {\it and\/}
absorbing {\MgII} halo gas kinematics has been performed for six
$z\sim 0.6$ highly inclined galaxies \citep{steidel02,ellison03}.
\citet{ellison03} found that the systemic velocity of a galaxy
coincided with the center of the absorption system, which spanned more
than 100~km~s$^{-1}$ about the systemic velocity. \citet{bond01a} used
expanding shell models to explain that this peculiar absorption
profile is likely caused by expanding supernovae--driven superbubbles.
\citet{steidel02} found that, in four of the five cases, the
velocities of all of the absorption components lie to one side of the
galaxy systemic redshift. The fifth case had a narrow, weak absorption
centered at galaxy systemic velocity.  Since the halo gas velocities
align in the same sense as the galaxy rotation, the velocity offsets
of the absorbing gas relative to the galaxy systemic velocity strongly
suggest ``disk--like'' rotation of the halo gas.  Using simple disk
halo models, \citet{steidel02} concluded that 
an extension of the disk rotation with a lagging halo component
(based upon properties of local galaxies' halo gas kinematics) 
was able to explain some of the gas kinematics. However, the
models were not able to account for the full velocity spreads of the
gas.


From a theoretical stand point, semi--analytical models and isolated
galaxy simulations
\citep[e.g.,][]{mo96,burkert00,lin00,maller04,chen08,kaufmann08,tinker08}
have been invoked to study isolated galaxy halos.  In these models,
{\MgII} absorption arises from condensed, infalling, pressure confined
gas clouds within the cooling radius of a hot halo. These models are
quite successful at reproducing the general statistical properties of
the absorber population. However, they lack the important dynamic
influences of the cosmic structure and local environments.

$\Lambda$CDM simulations have been able to synthesize the formation
and evolution galaxies within large scale structures. Recently,
\citet{ceverino09}, were able to naturally create, without ad--hoc
recipes, extended galactic scale outflows and metal enriched
multi--phased galactic gas halos. This was accomplished by studying
the detailed physics of the formation and evolution of the
multi--phase ISM in parsec resolution simulations.  These same
prescriptions were then successfully applied in their large scale
cosmological simulations.


Since these cosmological simulations include all the potential
structures that can influence halo gas dynamics and include the local
environment, they provide a promising technique for understanding the
role of gas in galaxy evolution. The quasar absorption line method can
be applied to simulations to examine structures selected by species
such as {\MgII}, in the vicinity of galaxies. The goal is to compare
directly observed absorbing halo gas kinematics and host galaxy
kinematics to those extracted from the simulations.  In order to
arrive at a deeper understanding, the observations should target
redshifts where detailed high quality kinematics can be obtained for a
sample of galaxies with a wide range of orientations with respect to
the quasar line of sight.

We have obtained ESI/Keck rotation curves of 10 intermediate redshift ($0.3 \leq z \leq 1.0$) galaxies for which we have high resolution HIRES/Keck or UVES/VLT quasar absorption profiles of {\MgII}, as well as WFPC--2/{\it HST\/} images.  In this paper we perform a kinematic comparison of 10 galaxies and their associated halo {\MgII} absorption. We define halo gas to be metal enriched structures that give rise to {\MgII} absorption such as extraplanar gas, outflows, tidal streams, filaments, and satellite galaxies.  We compare our observations with a simple rotating thick disk halo model \citep[similar to the one employed by][]{steidel02} and with the cosmological simulations of \citet{ceverino09}.  

The paper is organized as follows: In \S~\ref{sec:datakine}, we present our sample, and explain the data reduction and analysis. In \S~\ref{fields}, we present the results of our galaxy--{\MgII} absorption kinematic observations, and in \S~\ref{halo}, we compare the observed absorption velocities with a simple disk kinematic halo model.  In \S~\ref{sec:sim}, we discuss the details of the cosmological simulations. We study a simulated galaxy and its halo structures in detail. We analyze the integrated total hydrogen and {\HI} column density maps and the absorbing gas velocity distributions. We use these results to infer possible structures and kinematics drivers of that produce the observed {\MgII} absorption profiles.  We also compute the star formation rate and star formation surface density of the simulated galaxy and compare them to previous observational results, and with two of the galaxies in our sample. We end with our conclusions in \S~\ref{sec:conclusionkine}.  Throughout we adopt a $h=0.70$, $\Omega_{\rm M}=0.3$, $\Omega_{\Lambda}=0.7$ cosmology.



\section{Data and Analysis}
\label{sec:datakine}

\subsection{Sample Selection}

The selection of the sample presented in the study is based upon three steps.  (1) We compiled a list of {\MgII} absorbers in high resolution ($R = 45,000, \hbox{\rm FWHM} \simeq 6$~{\kms}, with $EW_r(2796)\geq 0.02$~\AA) HIRES/Keck \citep{vogt94} or UVES/VLT \citep{dekker00} quasar spectra.  We make no cut to the sample based upon equivalent width.  (2) We then compiled all subsequent deep ground based imaging and spectroscopic redshift surveys of the quasar fields and selected the galaxies that have {\it confirmed\/} redshifts aligned with {\MgII} absorption \citep[][ This paper]{bb91,bergeron92,sdp94,lowenthal95,gb97,chen98,lane98}.  (3) Finally, we selected galaxies for which WFPC--2/{\it HST\/} images were available and from which we can extract detailed galaxy morphological parameters \citep[see][]{kacprzak07}.  The final sample of ten galaxies have an impact parameter range of $26\leq D \leq 108$~kpc.

Our goal is to study the relationship between the spatial and kinematics relations between a galaxy and the {\MgII} absorbing gas in its vicinity.  In our sample, two of the {\MgII} absorbers appear to be associated with galaxies of similar luminosity that exhibit signs of interaction in the {\it HST\/} images.   These galaxies are G1 and G2 in the field of Q0450--132 and G1 and G2 in the field of Q1127--145 (see Figures~\ref{fig:0450_0454}$a,b$ and \ref{fig:1127}, respectively).   The Q0450--132 galaxies show tidal asymmetries and have projected separated of 12 kpc.  There are no additional candidate galaxies in the image within 100 kpc (projected) of the quasar.   The Q1127--145 galaxies also exhibit tidal asymmetries.  Their projected separation is roughly 50 kpc.  In the Q1127--145 field there is a third galaxy, G4, with much smaller luminosity, with a redshift that places it within 70~{\kms} of G2 and 25~{\kms} of G1.  

These particular systems, which are characterized by two roughly comparable luminosity galaxies with signs of interacting, pose an interesting challenge.  They indicate that some {\MgII} absorption is arising in the complex environment of a major--major galaxy interaction.  It is probably a fair statement to assert that interacting galaxies of roughly equal luminosity (and that is an important point) will have local environments very different than those of galaxies that clearly have no companion of comparable luminosity.  It becomes an intractable problem to discern what portion of the absorption may be arising with gas associated with one or the other galaxy in such a pair.  In cases where a single galaxy candidate can be assigned as the luminous host of the {\MgII}, it is possible to unambiguously study the spatial and kinematic relationships.  Of course, it is always possible a very low luminosity counterpart is below the detection of the images; but such a companion would indicate a minor--major interaction and not a major--major interaction like the Q0450-132 and Q1127--145 pairs.  Minor--major interactions would be more akin to the Magellanic galaxies in the 50 kpc vicinity of the Milky Way; they can be considered part of the Milky Way halo.  Such a distinction would equally apply to the single galaxies in this sample if there is an unseen minor companion.

In order to keep our sample as uniform as possible for the study of the spatial and kinematic connections between a galaxy and its associated absorbing gas, we limit our analysis to the single galaxy candidates when ambiguities would arise, such as comparison with galaxy inclination, position angle, and impact parameter.  There is a very different nature to the major--major interacting pairs toward Q0450-132 and Q1127--145 in that they may have a common gas envelope and therefore provide a slightly different probe of 
absorption galaxy properties.   In cases where ambiguities do not arise in the analysis, we include all the galaxies in our sample.   


\subsection{Quasar Spectroscopy}
\label{sec:qso_spec}

Details of the HIRES/Keck and UVES/VLT quasar observations are
presented in Table~\ref{tab:qsospec}. The HIRES spectra (except for
Q$0836+113$) were reduced using IRAF\footnote{IRAF is written and
supported by the IRAF programming group at the National Optical
Astronomy Observatories (NOAO) in Tucson, Arizona. NOAO is operated by
the Association of Universities for Research in Astronomy (AURA),
Inc.\ under cooperative agreement with the National Science
Foundation.}. The spectrum of Q$0836+113$ was reduced using the
MAKEE\footnote{http://spider.ipac.caltech.edu/staff/tab/makee}
package.  The UVES spectra were reduced using the standard ESO
pipeline and a custom code called the UVES Post--Pipeline Echelle
Reduction \citep[{\sc uves popler},][]{popler}. The quasar spectra are
both vacuum and heliocentric velocity corrected.  Analysis of the
{\MgII} absorption profiles was performed using graphic--based
interactive software of our own design
\citep[see][]{weakI,archiveI,cv01} for local continuum fitting,
objective feature identification, and measuring absorption
properties. The redshift for each {\MgII} system is computed from the
optical depth weighted mean of the absorption profile
\citep[see][]{cv01}. The typical absorption redshift uncertainty is
$\sim0.3$~{\kms}. The {\MgII} $\lambda 2796$ rest--frame equivalent
widths are adopted from \citet{kacprzak07}.  Velocity widths of
absorption systems are measured between the pixels where the
equivalent width per resolution element recovers to the $1~\sigma$
detection threshold \citep{weakI}.

\begin{deluxetable}{lcccr}
\tabletypesize{\scriptsize}
\tablecaption{Keck $+$ VLT Quasar Observations\label{tab:qsospec}}
\tablecolumns{5}
\tablewidth{0pt}

\tablehead{
\colhead{ }&
\colhead{ } &
\colhead{ } &
\colhead{ } &
\colhead{Exposure} \\
\colhead{QSO Field}&
\colhead{$z_{em}$} &
\colhead{Instrument} &
\colhead{Date (UT)} &
\colhead{(sec.)} 
}
\startdata
Q$0002+051$ &1.90  &HIRES & 1994 Jul. 05  & 2700\\
Q$0229+131$ &2.06  &HIRES & 1999 Feb. 08  & 3600\\
Q$0450-132$ &2.25  &HIRES & 1995 Jan. 24  & 5400\\
Q$0454-220$ &0.53  &HIRES & 1995 Jan. 22  & 5400\\
Q$0836+113$\tablenotemark{ b} &2.70  &HIRES & 1998 Feb. 26  & 5400\\
Q$1127-145$ &1.18  &UVES & $ \cdots $\tablenotemark{ a}  & 24,900\\
Q$2206-199$ &2.56  &UVES & $ \cdots $\tablenotemark{ a}  & 53,503
\enddata

\tablenotetext{a}{The Q$1127-145$ quasar spectrum was obtained over
multiple nights. The PIDs for this quasar are 67.A-0567(A) and
69.A-0371(A). The Q$2206-199$ quasar spectrum was also obtained over
multiple nights for the following PIDs 65.O-0158(A), 072.A-0346(A),
and 074.A-0201(A).}

\tablenotetext{b}{Data provided by Jason X. Prochaska \citep{prochaska07}.}
\end{deluxetable}

\subsection{HST Imaging and Galaxy Properties}


\begin{deluxetable}{lccl}[b]
\tabletypesize{\scriptsize}
\tablecaption{ WFPC--2/{\it HST\/} Observations\label{tab:HST}}
\tablecolumns{4}
\tablewidth{0pt}

\tablehead{
\colhead{ }&
\colhead{ } &
\colhead{Exposure} &
\colhead{ }\\
\colhead{QSO Field}&
\colhead{Filter} &
\colhead{(sec.)} &
\colhead{PID/PI}
}
\startdata
Q$0002+051$ & F702W  &4600 &  5984/Steidel \\
Q$0229+131$ & F702W  &5000 &  6557/Steidel \\
Q$0450-132$ & F702W  &2500 &  5984/Steidel \\
Q$0454-220$ & F702W  &1200 &  5098/Burbidge \\
Q$0836+113$ & F702W  &5000 &  6557/Steidel \\
Q$1127-145$ & F814W  &4400 &  9173/Bechtold  \\
Q$2206-199$ & F702W  &5000 &  6557/Steidel  
\enddata
\end{deluxetable}

All WFPC--2/{\it HST\/} images were reduced using the WFPC--2
Associations Science Products Pipeline (WASPP\footnote{Developed by
the Canadian Astronomy Data Centre (CADC) and the Space
Telescope--European Coordinating Facility (ST--ECF): {\it
http://archive.stsci.edu/hst/wfpc2/pipeline.html}}).  The WFPC--2
astrometry is calibrated to the USNO2 Catalog \citep{usno2}.  WASPP
data quality verifications include photometric and astrometric
accuracy and correctly set zero--points. Details of the WFPC--2/{\it
HST\/} observations are presented in Table~\ref{tab:HST}.  Galaxy
apparent and absolute magnitudes are adopted from
\citet{kacprzak07,kacprzak08}, respectively.  The $m_{F702W}$ and
$m_{F814W}$ magnitudes are based upon the Vega system. As described in
\citet{kacprzak07}, we used GIM2D \citep{simard02} to model the galaxy
morphologies, and measured the quasar--galaxy impact parameters,
galaxy sky orientations, inclination angles ($i$), and position angles
($PA$) of their major axes with respect to the quasar line of sight.
We fit each galaxy surface brightness profile with a Sersic bulge
component (for $0.2 \leq n \leq 4.0$) and an exponential disk
component. Additional modeled galaxy morphological parameters will be
presented elsewhere (Kacprzak {\etal} 2010, in preparation).


\begin{deluxetable}{lccrc}
\tabletypesize{\scriptsize}
\tablecaption{Keck--II/ESI Observations\label{tab:ESI}}
\tablecolumns{5}
\tablewidth{0pt}

\tablehead{
\colhead{ }&
\colhead{ } &
\colhead{ } &
\colhead{Exposure} &
\colhead{Slit}\\
\colhead{QSO Field}&
\colhead{z$_{abs}$} &
\colhead{Date (UT)} &
\colhead{(sec.)} &
\colhead{PA}
}
\startdata

Q$0002+051$ G1 &0.851407 & 2001 Oct. 16  & 7200 & $-8.5$ \\
Q$0229+131$ G1 &0.417337 & 2006 Dec. 24  & 6500 & 134 \\
Q$0450-132$ G1 &0.493937 & 2006 Dec. 24  & 5300 & $-13$ \\
Q$0450-132$ G2 &0.493937 & 2006 Dec. 24  & 5300 & $-13$ \\
Q$0454-220$ G1 &0.483338 & 2006 Dec. 24  & 4800 & 276 \\
Q$0836+113$ G1 &0.786726 & 2006 Dec. 24  & 5300 & 130 \\
Q$0836+113$ G2 &$\cdots$\tablenotemark{ a}& 2006 Dec. 24  & 5300 & 130 \\
Q$1127-145$ G1 &0.312710 & 2006 Dec. 24  & 3900 & 129  \\
Q$1127-145$ G2 &0.312710 & 2006 Dec. 24  & 4200 & 87.5 \\
Q$1127-145$ G3 &0.328266 & 2006 Dec. 24  &  600 & 87.5 \\
Q$2206-199$ G1 &1.017040 & 2001 Oct. 16  & 1800 & 75  
\enddata

\tablenotetext{b}{ There is no {\MgII} absorption associated with
  Q$0836+113$ G2 in the literature. Our HIRES data do not provide the
  necessary wavelength coverage.}

\end{deluxetable}

\begin{figure}
\includegraphics[angle=0,scale=0.47]{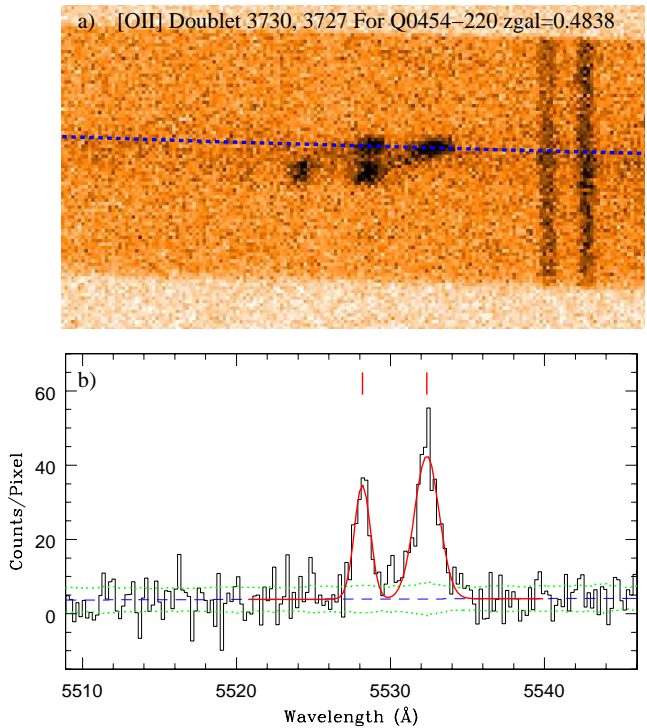}
\caption[angle=90]{(a) A 2D spectral region around the {\OII} doublet from the $z=0.4838$ galaxy in the quasar field Q$0454-220$. The {\OII} is spatially resolved and extends roughly $5''$ on the sky. The two sky lines extend the length of the $20''$ slit. --- (b) A 1D extraction of the above 2D image (thick dashed line in top panel) summed over 3 pixels in the spatial direction. The galaxy continuum fit is indicated by the dashed line. The 1~$\sigma$ uncertainty in the continuum is shown by the dotted lines bracketing the continuum.  The sky lines and sky signal have been subtracted out. The solid line shows the Gaussian fit to the emission lines and the tick marks indicates the centroids of the line.}
\label{fig:OII}
\end{figure}

\begin{deluxetable}{lcllr}
\tabletypesize{\footnotesize}
\tablecaption{$\hbox{{\rm Mg}\kern 0.1em{\sc ii}}$ Absorption And Galaxy Redshifts\label{tab:z}}
\tablecolumns{5}
\tablewidth{0pt}

\tablehead{
\colhead{ }&
\colhead{ } &
\colhead{ } &
\colhead{} &
\colhead{$\Delta v_{r}$\tablenotemark{ a}} \\
\colhead{QSO Field}&
\colhead{$z_{abs}$} &
\colhead{$z_{gal}$} &
\colhead{$\sigma z_{gal}$} &
\colhead{(km/s)}
}
\startdata
Q$0002+051$ G1 &0.851407 & 0.85180  & 0.000066 & $-66$ \\
Q$0229+131$ G1 &0.417337 & 0.4167   & 0.00020  & $+135$ \\
Q$0450-132$ G1 &0.493937 & 0.4941   & 0.00015  & $-33$ \\
Q$0450-132$ G2 &0.493937 & 0.4931   & 0.00012  & $+168$ \\
Q$0454-220$ G1 &0.483338 & 0.48382  & 0.000066 & $-98$ \\
Q$0836+113$ G1 &0.786726 & 0.78682  & 0.000028 & $-16$ \\
Q$1127-145$ G1 &0.312710 & 0.3132   & 0.00020  & $-112$ \\
Q$1127-145$ G2 &0.312710 & 0.3124   & 0.00013  & $+71$ \\
Q$1127-145$ G3 &0.328266 & 0.32847  & 0.000027 & $-46$ \\
Q$2206-199$ G1 &1.017040 & 1.01655  & 0.000013 & $+73$
\enddata

\tablenotetext{a}{$\Delta v_{r}$ is the rest--frame velocity offset
between the mean {\MgII} $\lambda 2976$ absorption line and the galaxy
where, $\Delta v_{r}=c(z_{abs}-z_{gal})/(1+z_{gal})$~\kms.}

\end{deluxetable}


\begin{deluxetable*}{lcllrrrc}[h]
\tabletypesize{\scriptsize}
\tablecaption{$\hbox{{\rm Mg}\kern 0.1em{\sc ii}}$ Absorption And Galaxy Redshift Field Survey\label{tab:zfield}}
\tablecolumns{8}
\tablewidth{0pt}

\tablehead{
\colhead{ }&
\colhead{Galaxy} &
\colhead{ } &
\colhead{$z_{gal}$} &
\colhead{ } &
\colhead{ } &
\colhead{ } &
\colhead{Galaxies in}\\
\colhead{QSO Field}&
\colhead{ID} &
\colhead{$z_{gal}$} &
\colhead{Reference\tablenotemark{ a}} &
\colhead{$D$ (kpc)} &
\colhead{$z_{abs}$} &
\colhead{$W_r(2796)$ {\AA}}&
\colhead{This Study}
}
\startdata
Q$0002+051$ & G1 & 0.85180  &  1  & $25.9\pm0.5$  & 0.851407 & $1.119\pm0.013$&X  \\
            & G2 & 0.592    &  2  & $36.2\pm0.4$  & 0.591365 & $0.102\pm0.002$&  \\
            & G3 & 0.298    &  2  & $59.3\pm0.3$  & 0.298059 & $0.246\pm0.004$&  \\
Q$0229+131$ & G1 & 0.4167   & 1,3 & $37.5\pm0.5$  & 0.417337 & $0.816\pm0.022$&X  \\
Q$0450-132$ & G1 & 0.4941   & 1,2 & $50.1\pm0.4$  & 0.493937 & $0.674\pm0.026$&X  \\
            & G2 & 0.4931   & 1,2 & $62.7\pm0.7$  & 0.493937 & $0.674\pm0.026$&X  \\
Q$0454-220$ & G1 & 0.48382  & 1,4 & $107.9\pm0.8$ & 0.483338 & $0.426\pm0.007$&X  \\
Q$0454-220$ & G2 & 0.3818   & 4   & $103.4\pm0.3$ & $\cdots$ & $<0.02$~$(3~\sigma)$& \\
Q$0836+113$ & G1 & 0.78682  & 1,5 & $26.9\pm0.9$  & 0.786726 & $2.148\pm0.023$&X  \\
            & G2 & 0.48288  & 1   & $29.1\pm0.3$ & $\cdots$ & $\cdots$\tablenotemark{ b}&\\
Q$1127-145$ & G1 & 0.3132   & 1,3 & $45.6\pm0.3$  & 0.312710 & $1.773\pm0.006$&X   \\
            & G2 & 0.3124   & 1,3 & $81.0\pm0.3$  & 0.312710 & $1.773\pm0.006$&X   \\
            & G3 & 0.32847  & 1   & $91.4\pm0.2$  & 0.328266 & $0.029\pm0.003$&X   \\
            & G4 & 0.3121   & 6   & $18.2\pm0.3$  & 0.312710 & $1.773\pm0.006$&  \\
Q$2206-199$ & G1 & 1.01655  & 1,7 & $104.6\pm1.4$ & 1.017040 & $1.057\pm0.005$&X \\
            & G2 & 0.948    & 7   & $87.2\pm0.5$  & 0.948361 & $0.253\pm0.002$& \\
            & G3\tablenotemark{ c} & 0.755    & 8   & $44.2\pm0.7$  & 0.751923 & $0.886\pm0.003$& \\
\enddata

\tablenotetext{a}{Galaxy Identification: (1) This paper,
  (2)~\citet{sdp94}, (3)~\citet{bb91}, (4)~\citet{chen98},
  (5)~\citet{lowenthal95}, (6)~\citet{lane98}, (7)~\citet{bergeron92},
  and (8)~\citet{gb97}. We list the redshift for galaxies that were
  derived for this work.}

\tablenotetext{b}{ There is no {\MgII} absorption associated with
  Q$0836+113$ G2 in the literature. Our HIRES data do not provide the
  necessary wavelength coverage.}

\tablenotetext{c}{ G3 was reported as a galaxy by \citet{gb97}.
However, our spectroscopic observations reveal that this unresolved
object is a Galactic star.}
\end{deluxetable*}

\subsection{Galaxy Spectroscopy}

The ESI/Keck \citep{sheinis02} galaxy spectra were obtained over two
nights; two were obtained in October 2001 and eight were obtained in
December 2006.  Details of the ESI/Keck observations are presented in
Table~\ref{tab:ESI}. For each galaxy, the slit position angle was
chosen to lie along the galaxy major axis (except for Q$450-132$ G2).
The slit length is $20''$. Thus in some cases, we were able to
simultaneously position two galaxies on a slit. Exposure times range
between $600-7200$s per galaxy. The wavelength coverage of ESI is 4000
to 10,000~{\AA}, which allows us to obtain multiple emission lines
(such as {\OII} doublet, $\rm{H}\beta$, {\OIII} doublet,
$\rm{H}\alpha$, [\NII] doublet, etc.)  with a velocity resolution of
$11$~\kms~pixel$^{-1}$ (${\rm FWHM}\sim45$~km/s).

In October 2001, the data were obtained with a $1''$ slit and
$2\times2$ binning.  The mean seeing was $0.7''$ (${\rm FWHM}$) with
partial cloud coverage.  In December 2006, the data were obtained
using a $0.75''$ slit with $2\times 1$ binning.  Binning by two in the
spatial directions results in pixel sizes of $0.27-0.34''$ over the
orders of interest. The mean seeing was $0.8''$ (${\rm FWHM}$) with
clear skies.

\begin{figure*}
\includegraphics[angle=0, scale=0.52]{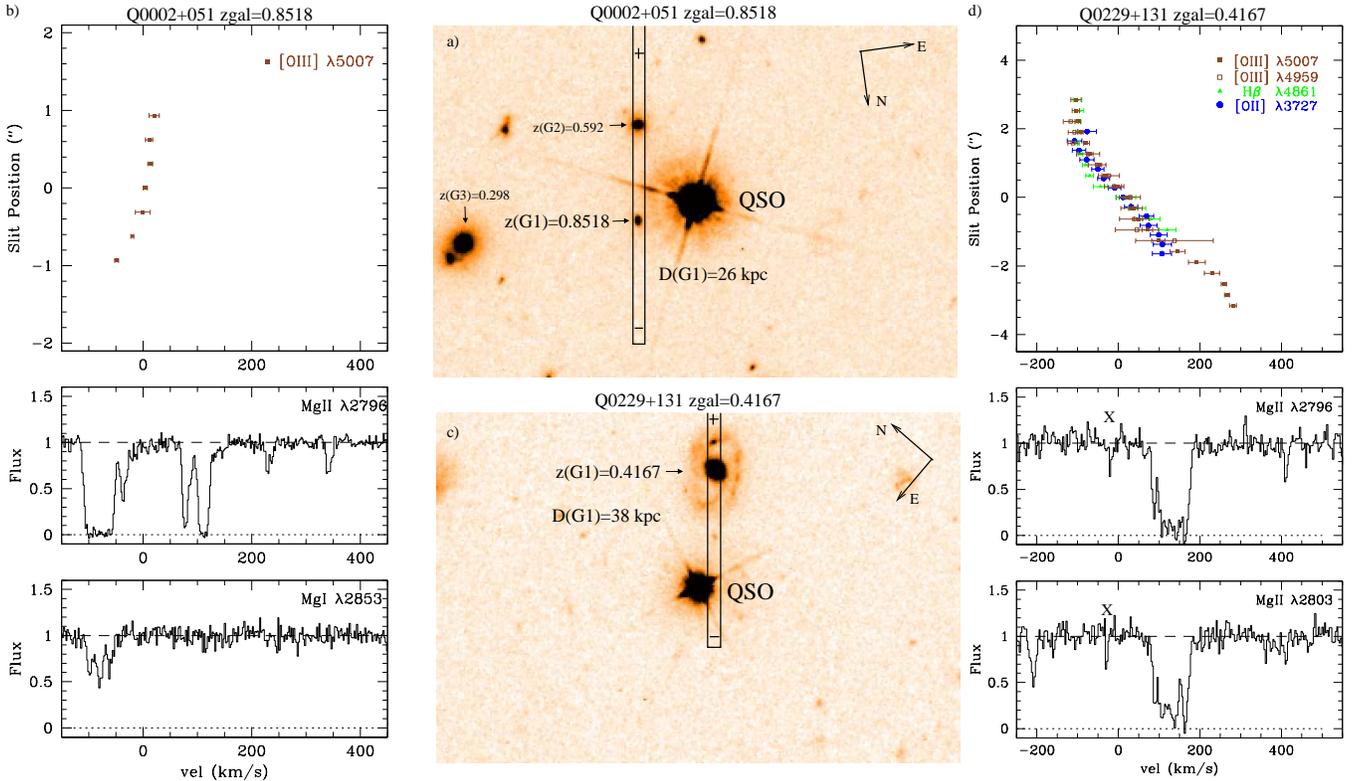}
\caption[angle=90]{(a) A $30'' \times 20''$ F702W WFPC--2/{\it HST\/}
  image of the quasar field Q$0002+051$.The ESI/Keck slit is
  superimposed on the image.  The '$+$' and '$-$' on the slit indicate
  the positive and negative arcseconds where $0''$ is defined at the
  target galaxy center. The targeted galaxy is a compact galaxy with
  $z=0.8518$ at impact parameter $D=25.9$~kpc. Two other galaxies have
  been identified in this field and their redshifts are indicated. ---
  (b) The $z=0.8518$ galaxy rotation curve and the HIRES/Keck
  absorption profiles aligned with the galaxy systemic velocity.  ---
  (c) Same as (a) except for the Q$0229+131$ quasar field. The
  targeted galaxy is a low inclination spiral with $z=0.4167$ at
  $D=37.5$~kpc.--- (d) Same as (b) except the $z=0.4167$ galaxy in the
  Q$0229+131$ field ({\MgI} $\lambda 2853$ absorption is not shown
  because it is blended with {\SiIV} $\lambda 1394$ associated with a
  $z=1.9024$ {\CIV} absorber). Note that the {\MgII} absorption
     resides fully to one side of the galaxy systemic velocity
    and also aligns with one arm of the rotation curve.}
\label{fig:0002_0229}
\end{figure*}

The data were reduced using the standard echelle package in IRAF.  We used internal quartz illumination flat fields to eliminate pixel to pixel variations.  In the science frames, sky subtraction was performed by fitting a polynomial function to each spatial column. A quasar or bright star spectrum in the same field was obtained and used as a trace in order to facilitate the extraction of the galaxy spectrum.  A spatially integrated spectrum was extracted in order to obtain an accurate galaxy redshift from the centroids of multiple emission lines. The lines listed in the legends of Figures~\ref{fig:0002_0229}--\ref{fig:2206} were used to determine the galaxy redshifts. The rest--frame vacuum wavelength used for each emission line was obtained from the National Institute of Standards and Technology (NIST) database.  Each spectrum was wavelength calibrated using CuArXe arc line lamps. Spectra were calibrated in IRAF using standard stars taken during the
night of the observation.  The flux is accurate to $\sim$10\% and we
have made no corrections for slit loss, or Galactic reddening.

In Figure~\ref{fig:OII}, we show an example of a 2D spectrum of the spatially resolved {\OII} doublet from the $z=0.4838$ galaxy in the Q$0454-220$ field.  We used a Gaussian fitting algorithm \citep[see][]{archiveI}, which computes best fit Gaussian amplitudes, widths, and centers (redshift), to the galaxy emission and absorption lines.  The galaxy redshift was computed from the mean redshift of all the detected lines. Emission lines and absorption lines used to calculate the galaxy redshift must have been detected at the $3~\sigma$ level.  The adopted redshift uncertainty for each galaxy was computed from the standard deviation in the redshifts computed from each emission line. The galaxy redshifts are listed in Table~\ref{tab:z}; their accuracy ranges from 2--45~\kms. The galaxy velocity offsets from the optical depth weighted mean {\MgII} absorption are also listed in Table~\ref{tab:z} and range from $-12$ to $+168$~{\kms}.


The rotation curve extraction was performed following the methods of \citet{vogt96} and \citet{steidel02}. We extract individual one--dimensional spectra by summing three--pixel wide apertures (corresponding to approximately one resolution element of $0.82-1.01''$) at one pixel spatial increments along the slit.  An error spectrum is also extracted for each of these apertures.  To obtain accurate wavelength calibrations, we extracted spectra of CuArXe arc line lamps at the same spatial pixels as the extracted galaxy spectra.  Fitted arc lamp exposures (CuArXe) provided a dispersion solution accurate to $\sim0.15$~\AA, or about $6.5$~{\kms} at the wavelengths of interest. Galaxy spectra are both vacuum and heliocentric velocity corrected for comparison with the absorption--line kinematics.  Each galaxy emission line (or absorption line in some cases) was fit with a single Gaussian (except the {\OII} doublet was fit with a double Gaussian) in order to extract the wavelength centroid for each emission line.  

An example of a three--pixel wide spectral extraction from the 2D spectrum is shown in Figure~\ref{fig:OII}$b$ where the {\OII} $\lambda3727$ and $\lambda3730$ lines are detected at the $22\sigma$ and $17\sigma$ level, respectively (the significance level is the ratio of the measured equivalent width to the uncertainty in the equivalent width based upon error propagation using the error spectrum extracted for the same three--pixel aperture).  The dashed line in Figure~\ref{fig:OII}$a$ provides the spatial cut for which the spectrum in Figure~\ref{fig:OII}$b$ is illustrated.  The velocity offsets for each emission line in each extraction were computed with respect to the redshift zero point determined for the galaxy (Table~\ref{tab:z}). The rotation curves for the 10 galaxies obtained with ESI/Keck are presented in Figures~\ref{fig:0002_0229}--\ref{fig:2206}.



\section{Discussion of Individual Fields}\label{fields}

Here we discuss the halo gas and galaxy kinematics of ten galaxies in
seven different quasar fields. In Table~\ref{tab:zfield}, we list all
the galaxies in each field that have spectroscopically confirmed
redshifts.  The table columns are (1) the quasar field, (2) the galaxy
ID, (3) the galaxy redshift, (4) the reference(s) for the galaxy
identification, (5) the quasar--galaxy impact parameter, $D$, and
uncertainty, (6) the {\MgII} absorption redshift, and (7) the
rest--frame {\MgII} $\lambda 2796$ equivalent width, $W_r(2796)$, and
uncertainty. Three new galaxies (Q$0002+051$ G1, Q$0836+133$ G2, and
Q$1127-145$ G3) have been spectroscopically identified in this work.

Most of the quasar fields listed in Table~\ref{tab:zfield} have been spectroscopically surveyed for all galaxies with $m_r\lesssim23$, which translates to $L_{\star}\simeq 0.08L_{\star}$ for $z \simeq 0.5$, out to $\sim1'$ \citep[e.g.,][]{sdp94,gb97,chen98}.  These works have been instrumental in developing our current picture of the galaxy--absorber connection at intermediate redshifts.  It remains possible that additional galaxies below this luminosity lurk in front of the quasar and may be associated with the absorbing gas.  It is also possible that a less than 100\% completeness in the confirmation of galaxy redshifts may result in an additional galaxy or galaxies also associated with the absorption.  The concerns and caveats associated with incorrect idenitifications and the conclusion drawn for those works therefore also apply to this study. Galaxies fainter than $m_r\sim23$, or low mass galaxies hidden by the quasar PSF, such as satellites or minor companions, can be considered part of the halo of the large host galaxy.  

In the following subsections, we discuss only the ten galaxies
selected for this study (see column 8 of Table~\ref{tab:zfield}).
Detailed images of the galaxies are presented in \citet{kacprzak07}.
A summary of the impact parameters, galaxy maximum rotation
velocities, and GIM2D model inclinations and position angles
($PA\equiv$ angle between the galaxy major axis and the quasar line of
sight) is listed in Table~\ref{tab:params}.  Galaxy redshifts will
only be quoted to four significant figures from here on for
simplicity.  We will later discuss kinematic halo models in
\S~\ref{halo}.

\subsection{Q0002+051 G1}
A WFPC--2/{\it HST\/} image of the Q$0002+051$ field is shown is
Figure~\ref{fig:0002_0229}$a$.  The $z=0.8518$ galaxy, G1, was
targeted for this study. It was first assumed by \citet{sdp94} to be
the absorbing galaxy associated with the {\MgII} absorption at
$z=0.851407$ given its colors and its proximity to the quasar. We
report the first redshift confirmation of this galaxy. The galaxy
redshift was identified by an {\OIII} and a weak {\OII} emission line
(the {\OII} line was used only for the determination of the galaxy
redshift and provided no kinematic information).  In
Figure~\ref{fig:0002_0229}$a$, we note that the ESI/Keck slit was
placed across both G1 and G2 (where '$+$' and '$-$' on the slit
indicate the positive and negative arcseconds relative to the center
of the galaxy, respectively). We did not detect any identifiable
emission lines from the galaxy G2.

The {\MgII} absorption was first reported by \citet{bechtold84} and
then confirmed by \citet{ss92}. The HIRES/Keck spectrum was originally
presented by \citet{csv96} and has $W_r(2796)=1.119$~{\AA}.

The G1 galaxy has a compact morphology. It has an absolute $B-$band
magnitude of $M_B=-21.3$ which translates to a luminosity
$L_B=0.92L_B^\ast$.  The galaxy has inclination angle $i=38^\circ$ and
$PA=43^\circ$.  G1 is at a projected distance of $D=25.9$~kpc from the
quasar line of sight.  In Figure~\ref{fig:0002_0229}$b$, the spatial
radial velocity, as derived from {\OIII}, is shown.  It appears G1 is
not predominantly rotating.  The projected velocity shear is
$49$~{\kms}.  Displayed in the lower panels are the {\MgII} and {\MgI}
absorption profiles, shown on the same velocity scale. The mean
optical depth absorption redshift is offset by $-66$~{\kms} from the
galaxy systemic velocity. This {\MgII} absorption system is the most
kinematically complex absorption profile in our sample. Four
absorption sub--systems are spread out over a total velocity of
$\sim475$~{\kms}.  The blue--shifted extreme wing of G1's velocity
shear is partially consistent with the velocities of the dominant
saturated component of the {\MgII}.  However, it is inconsistent with
the {\MgII} gas at large positive velocities.  The {\MgI} gas, which
may trace more neutral and higher column density gas, has velocities
is also inconsistent with the velocities of G1.

\begin{figure*}
\includegraphics[angle=0, scale=0.52]{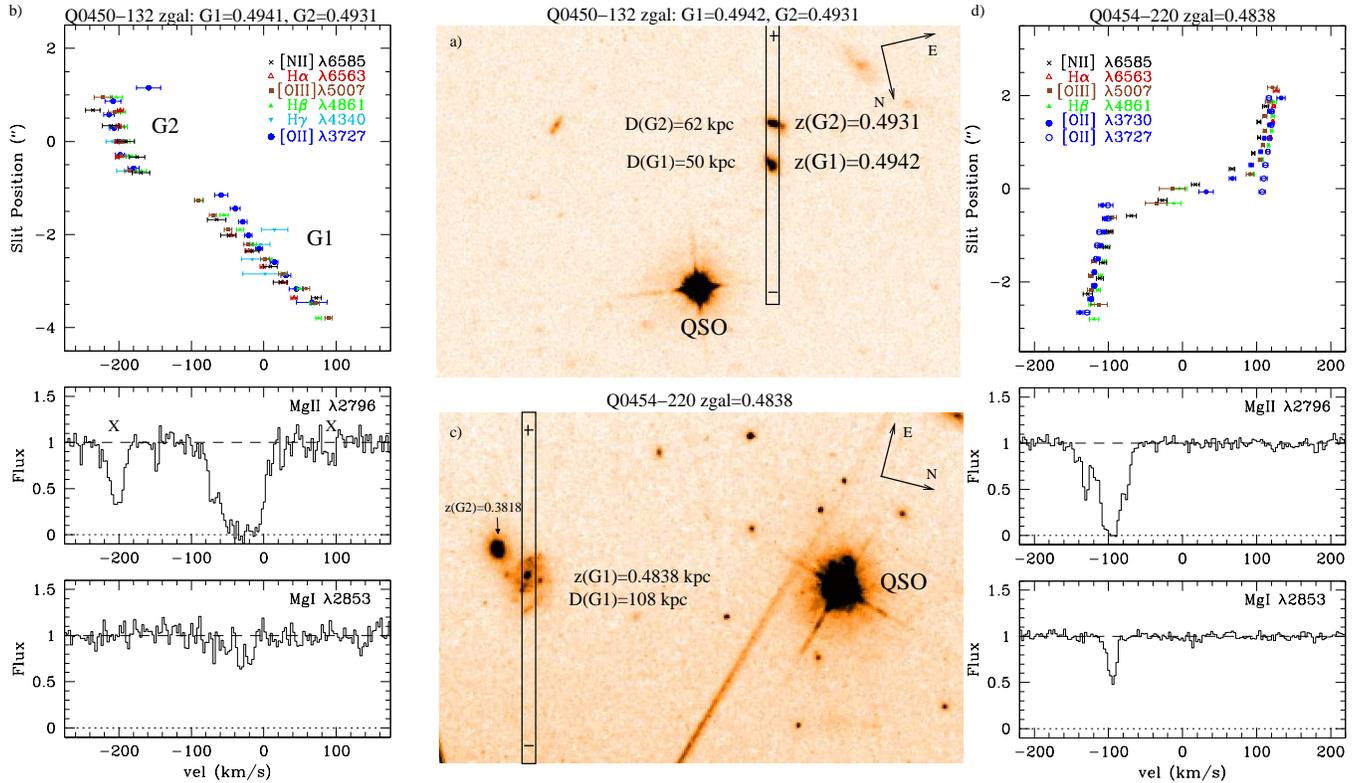}
\caption[angle=90]{(a) Same as Figure~\ref{fig:0002_0229}$a$ except
for the quasar field Q$0450-132$. The target spiral galaxies G1 and G2
are possibly interacting galaxies, indicated by the single--sided
tidal tails. G1 is at a redshift of $z=0.4942$ at $D=50.1$~kpc. G1 is
at a redshift of $z=0.4931$ at $D=62.7$~kpc. ---(b) Same
Figure~\ref{fig:0002_0229}$b$ except G1 and G2 in the Q$0450-132$
field. --- (c) Same as (a) except the $z=0.4838$ galaxy, G1, in the
Q$0454-220$ field. The targeted galaxy is a spiral galaxy at an impact
parameter $D=107.9$~kpc.  --- (d) Same as (b) except for the G1 galaxy
in the Q$0454-220$ field.}
\label{fig:0450_0454}
\end{figure*}

\begin{figure*}
\includegraphics[angle=0, scale=0.52]{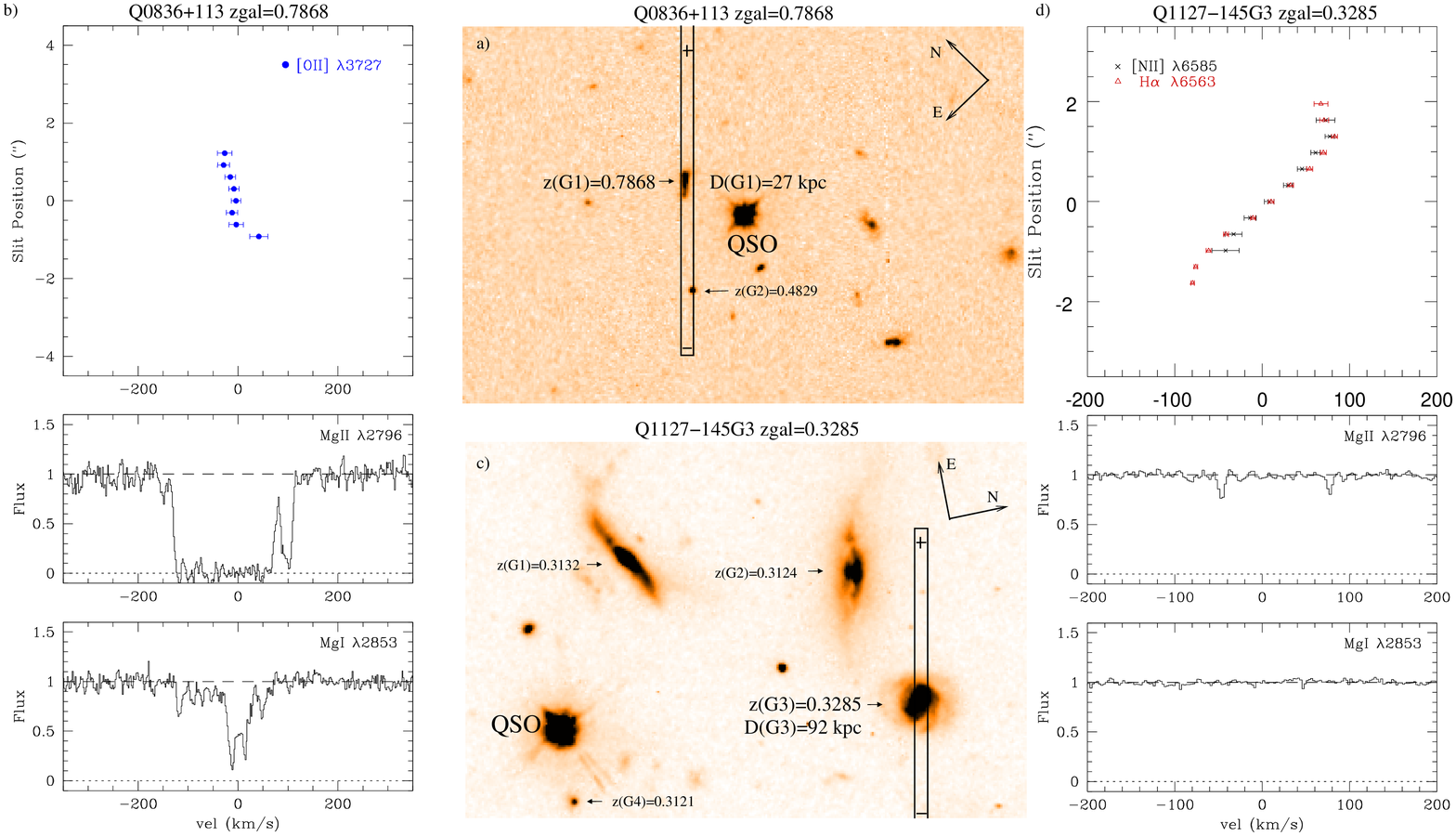}
\caption[angle=90]{(a) Same as Figure~\ref{fig:0002_0229}$a$ except
for the quasar field Q$0836+113$. The target spiral galaxy G1 is at a
redshift of $z=0.7868$ at $D=26.9$~kpc.  ---(b) Same
Figure~\ref{fig:0002_0229}$b$ except G1 in the Q$0836+113$ field.  ---
(c) Same as (a) except the $z=0.3285$ galaxy, G1, in the Q$1127-145$
field. The targeted galaxy is a spiral galaxy at an impact parameter
$D=91.4$~kpc.  --- (d) Same as (b) except for the G1 galaxy in the
Q$1127-145$ field.}
\label{fig:0836_1127}
\end{figure*}

\subsection{Q0229+131 G1}

A WFPC--2/{\it HST\/} image of the Q$0229+131$ field is shown in
Figure~\ref{fig:0002_0229}$c$. The $z=0.4167$ galaxy, G1, was
initially spectroscopically identified by \citet{bb91}. The first
detection of {\MgII} absorption at the galaxy redshift was reported by
\citet{sargent88}.  A HIRES/Keck spectrum of this absorber was first
presented by \citet{cwc-china} and has  $W_r(2796)=0.816$~{\AA}.

This moderately inclined, $i=58^\circ$, bulge dominated galaxy has two
large grand design spiral arms similar to a local Sb galaxy.  One of
the arms contains a bright {\HII} region. The galaxy position angle is
$PA=22^\circ$. G1 has an absolute $B-$band magnitude of $M_B=-20.8$
which translates to $L_B=0.74L_B^\ast$. G1 is at a projected distance of
$D=37.5$~kpc from the quasar line of sight.

In Figure~\ref{fig:0002_0229}$d$, the rotation curve, derived from
{\OII}, $\rm{H}\beta$, and {\OIII} emission lines, is shown. The
asymmetric rotation curve has a maximum observed velocity of
$280$~{\kms}. The asymmetry may be a result of the asymmetric spiral
arms. Below the rotation curve, the {\MgIIdblt} absorption profiles
are shown on the same velocity scale. The mean absorption redshift is
offset by $+135$~{\kms} from the galaxy systemic velocity.  The {\MgI}
profile is not shown here since it is completely bended with a four
component {\SiIV} $\lambda 1394$ complex from a $z=1.9024$ {\CIV}
absorber \citep{sargent88}.  This {\MgII} absorption system has a
velocity spread of roughly 112~{\kms} with a single cloud residing
roughly $+280$~{\kms} from the main component.  The main absorption
component aligns with the redshifted wing of the galaxy rotation curve
arising from the spiral arm nearest the quasar. The galaxy rotation
velocities are not consistent with the outlying cloud.

\subsection{Q0450-132 G1,G2}

A WFPC--2/{\it HST\/} image of the Q$0450-132$ field is shown in
Figure~\ref{fig:0450_0454}$a$. Two galaxies were placed along the same
slit. The redshifts of this double pair were initially obtained by
\citet{sdp94}. This absorption system was discovered during the survey
of \citet{sargent88} and the HIRES/Keck spectrum was first presented
by \citet{cwcthesis}. The {\MgII} absorption has an equivalent width
of $W_r(2796)=0.674$~{\AA}.

The galaxy closer to the quasar, G1, has a redshift of $z=0.4941$ and
is at an impact parameter of $D=50.1$~kpc. The galaxy has an
inclination of $i=66^\circ$ with a $PA=42^\circ$. G1 has an absolute
magnitude of $M_B=-19.7$ which translates to a $L_B=0.25L_B^\ast$
galaxy.  The galaxy further from the quasar, G2, has a redshift of
$z=0.4931$ and has an impact parameter of $62.7$~kpc. The galaxy has
an inclination of $i=75^\circ$ with a $PA=54^\circ$. G2 has an
absolute magnitude of $M_B=-19.7$ which translates to
$L_B=0.26L_B^\ast$.  The galaxies are spatially separated by only
$14$~kpc projected, and by a line of sight velocity of
$201$~\kms. Both galaxies appear to have single-sided tidal tails
and show strong morphological evidence of a previous interaction or
harassment.

The rotation curves of both galaxies, as derived from {\OII},
$\rm{H}\gamma$, $\rm{H}\beta$, {\OIII}, $\rm{H}\alpha$, and [\NII],
are presented in Figure~\ref{fig:0450_0454}$b$. The maximum observed
rotation velocities for G1 and G2 are $98$~{\kms} and $47$~{\kms},
respectively.  Below the rotation curve, the {\MgII} and {\MgI}
absorption profiles are shown on the same velocity scale. The mean
absorption redshift is offset by $-33$~{\kms} from the galaxy systemic
velocity of G1. The {\MgII} absorption is a single kinematic region
with a velocity width of $103$~{\kms}. The absorption aligns, in
velocity, with the blueshifted wing of the rotation curve of G1.  The
{\MgI} aligns more closely to the systemic velocity of G1. The galaxy
G2 does not have much rotation since the slit position angle is close
to its minor axis.
 
\subsection{Q0454-220 G1}

A WFPC--2/{\it HST\/} image of the Q$0454-220$ field is shown in
Figure~\ref{fig:0450_0454}$c$. The $z=0.4838$ galaxy, G1, was targeted
for this study. G1 was spectroscopically identified by
\citet{chen98}. The {\MgII} absorption was first reported by
\citet{begeron84} and then confirmed by \citet{tytler87}. The
HIRES/Keck spectrum was originally presented by \citet{cv01}.  The
{\MgII} equivalent width is $W_r(2796)=0.426$~{\AA}.

G1 is a spiral galaxy that has a perturbed morphology with one
extended spiral arm. It has a compact bulge and several bright {\HII}
regions. The galaxy has an inclination of $i=41^\circ$ and a position
angle of $PA=76^\circ$. It has an absolute $B-$band magnitude of
$M_B=-21.0$ which translates to $L_B=0.90L_B^\ast$. G1 is at a
projected distance of $D=107.9$~kpc from the quasar line of sight.

The rotation curve, as derived from {\OII}, $\rm{H}\beta$,
{\OIII}, $\rm{H}\alpha$, and [\NII], is presented in
Figure~\ref{fig:0450_0454}$d$.  The rotation curve flattens out at a
maximum observed velocity of $138$~{\kms}. Below the rotation curve,
the {\MgII} and the {\MgI} absorption profiles are shown on the same
velocity scale. The mean absorption redshift is offset by $-98$~{\kms}
from the galaxy systemic velocity. This {\MgII} absorption is a single
kinematic system having a velocity spread of roughly 93~{\kms}.  The
blue shifted component of the rotation curve has velocities that are
consistent with the {\MgII} and {\MgI} absorption profiles.

\subsection{Q0836+113 G1}

A WFPC--2/{\it HST\/} image of the Q$0836+113$ field is shown in
Figure~\ref{fig:0836_1127}$a$.  The $z=0.7868$ galaxy, G1, was
initially spectroscopically identified by \citet{lowenthal95}. The
first {\MgII} absorption detection at the galaxy redshift was reported
by \citet{turnshek89}.  The HIRES/Keck absorption profiles of this
absorber is first presented in this work
(Figure~\ref{fig:0836_1127}$b$) and has $W_r(2796)=2.15 \pm
0.02$~{\AA}.

The galaxy G1 appears to be an edge--on spiral with an asymmetric
brightness profile. The morphology is similar to that of the two
galaxies in Q$0450-132$, which have single--sided tidal tails. This
almost edge--on galaxy has an inclination of $i=78^\circ$ and
$PA=57^\circ$. G1 has an absolute $B-$band magnitude of $M_B=-20.5$
which translates to $L_B=0.39L_B^\ast$. The galaxy is at a projected
distance of $D=26.9$~kpc from the quasar.

The spatial radial velocity of G1, as derived from {\OII}, is
presented in Figure~\ref{fig:0836_1127}$b$.  The data suggest that G1
exhibits more of a global shear than rotation. The maximum observed
shear velocity is $\sim 42$~{\kms}. Below the velocity curve, the
{\MgII} and the {\MgI} absorption profiles are shown on the same
velocity scale. The mean absorption redshift is offset by $-16$~{\kms}
from the galaxy systemic velocity. The {\MgII} absorption is a single
kinematic component having a velocity spread of roughly
282~{\kms}. Most of the component is composed of highly saturated
clouds. From the {\MgI} profile, one can resolve the individual clouds
that are saturated in {\MgII}.  The galaxy velocities are
consistent with the velocities of the absorbing gas. However, there is
a large amount of {\MgII} gas that has greater velocities than those
of the galaxy.  The bulk of the more neutral/high column density gas,
as indicated by {\MgI}, is at the galaxy systemic velocity.

\begin{figure*}
\includegraphics[angle=0, scale=0.52]{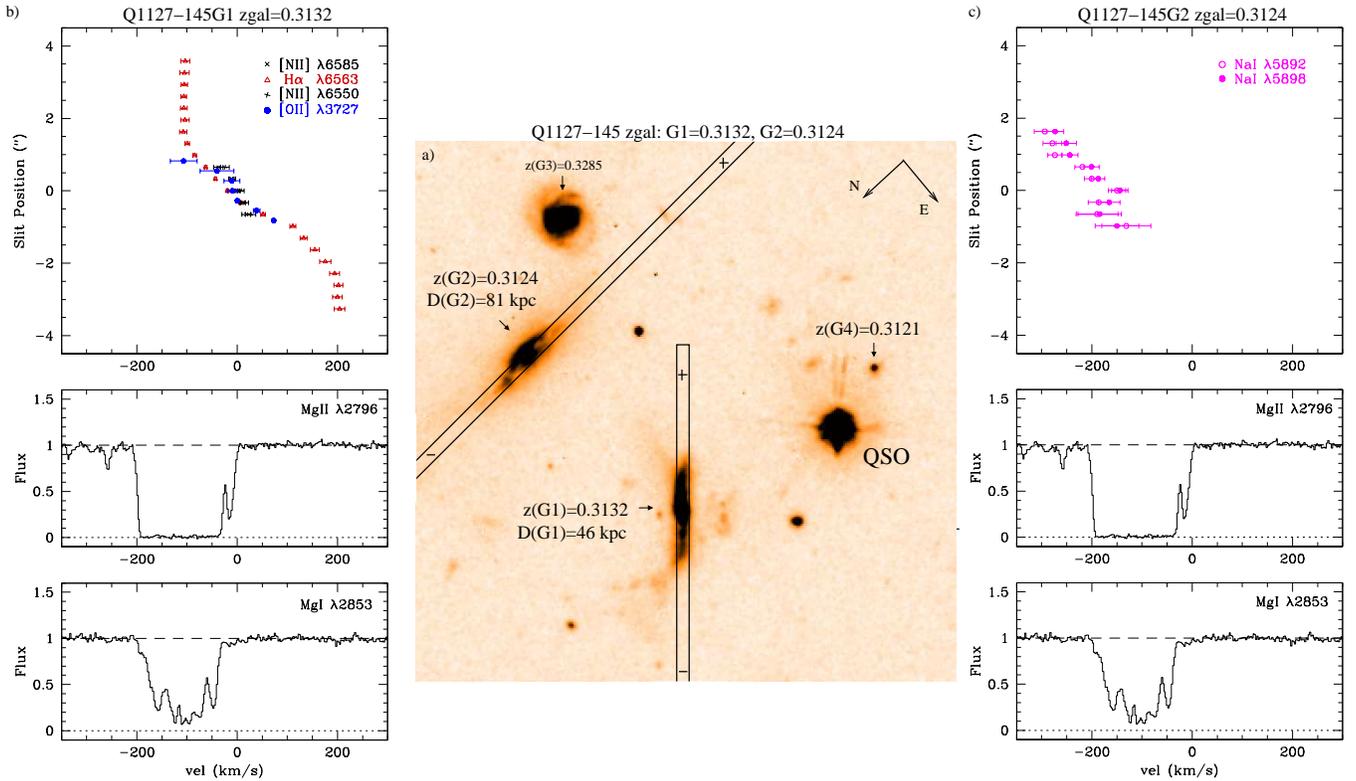}
\caption{ (a) Same as Figure~\ref{fig:0002_0229}$a$ except that this
  is a $30'' \times 30''$ F814W WFPC--2/{\it HST\/} image of the
  quasar field Q$1127-145$. The two targeted spiral galaxies G1 and G2
  have redshifts of $z=0.3132$ and $z=3124$, respectively. G1 and G2
  are at impact parameters of $D=45.6$~kpc and $D=81.0$~kpc,
  respectively. --- (b) Same as Figure~\ref{fig:0002_0229}$b$ except
  G1 in the Q$1127-145$ field.  --- (b) Same as
  Figure~\ref{fig:0002_0229}$b$ except G2 in the Q$1127-145$ field.}
\label{fig:1127}
\end{figure*}

\subsection{Q1127-145 G3}
A WFPC--2/{\it HST\/} image of the Q$1127-145$ field is shown in
Figure~\ref{fig:0836_1127}$c$.  We present a newly identified galaxy,
G3, at $z=0.3285$. The galaxy was identified by $\rm{H}\alpha$ and
[\NII] emission lines. The absorption was also recently discovered,
and is presented here in Figure~\ref{fig:0836_1127}$d$. This weak
system has an equivalent width of $W_r(2796)=0.029$~{\AA}.

The face--on galaxy has an inclination of $i=1^\circ$ with a
$PA\sim69^\circ$ (the $PA$ is highly uncertain since $i\simeq 0$). The
galaxy has a large bar with a sizable bright bulge; similar to a local
SBb galaxy. It has absolute magnitude of $M_B=-20.51$ which translates
to $L_B=0.60L_B^\ast$. The galaxy is at a projected distance of
$D=91.4$~kpc from the quasar.

The rotation curve of G3 is presented in
Figure~\ref{fig:0836_1127}$d$.  The maximum observed rotational
velocity is $\sim 80$~{\kms}. Given that significant rotation is
observed, the galaxy is most likely not completely face--on as the
GIM2D model inclination suggests. Below the rotation curve, the
{\MgII} and the {\MgI} absorption profiles are shown on the same
velocity scale. The absorption redshift is offset by $-46$~{\kms} from
the galaxy systemic velocity.  The {\MgII} absorption contains two
separate single cloud components. Both clouds have a velocity spread
of $\sim 15$~{\kms} and are separated by $125$~{\kms}.  No significant
{\MgI} ($3\sigma$, $W_r\leq0.003$~\AA) is detected. The projected
galaxy rotation velocities are consistent with both cloud velocities;
each cloud aligns with each side of the rotation curve.

\begin{figure*}
\begin{center}
\includegraphics[angle=0, scale=0.52]{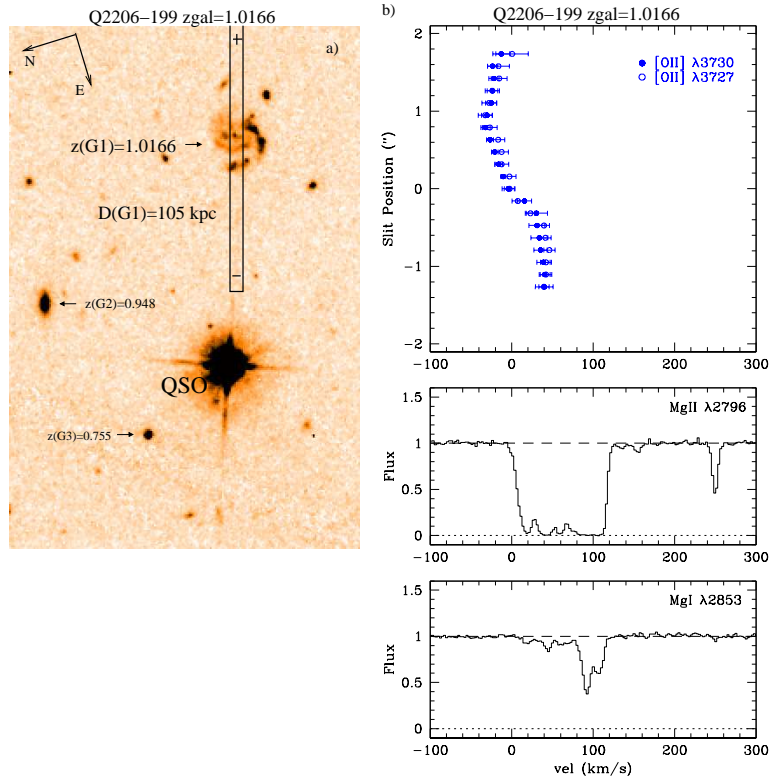}
\end{center}
\caption[angle=90]{(a) Same as Figure~\ref{fig:0002_0229}$a$ except
for the quasar field Q$2206-199$. The target spiral galaxy G1 is at a
redshift of $z=1.10166$ at $D=104.6$~kpc.  ---(b) same
Figure~\ref{fig:0002_0229}$b$ except G1 in the Q$2206-199$ field.}
\label{fig:2206}
\end{figure*}

\subsection{Q1127-145 G1,G2}\label{sec:1127}

The history of the Q$1127-145$ $z=0.312710$ absorption system is quite
complex.  The {\MgII} absorption was initially detected by
\citet{bb91} and was determined to be a DLA \citep[see][]{rao00} since
{\it HST\/} UV data show a damped Ly$\alpha$ with
$N_{HI}=5.1\times10^{21}$~cm$^{-2}$ \citep{lane98}.  The equivalent
width of the system is $W_r(2796)=1.773$~{\AA}.

The true identity of the absorbing galaxy has been a topic of debate
in the literature. A WFPC--2/{\it HST\/} image of the Q$1127-145$
field is shown in Figure~\ref{fig:1127}$a$.  \citet{bb91}
spectroscopically identified G1 and G2 to be at the redshift of the
absorption. G1 exhibits strong emission lines and G2 has no detectable
emission lines. G1 was assumed to be the absorbing galaxy, since it is
closer to the quasar and has significant star formation \citep{bb91}.
\citet{lane98} later spectroscopically identified G4 via an {\OIII}
doublet with $z=0.3121$, which was also consistent with the absorption
redshift and was assumed to be the absorber only due to its proximity
to the quasar. \citet{lane98} state that it is also possible that the
three galaxies may have undergone a strong interaction where the
absorption could arise from tidal debris.  In
Figure~\ref{fig:1127}$a$, it is apparent from the tidal disturbances,
that G1 has undergone interactions in the past. G2 also exhibits some
tidal material to the north of the galaxy (which is less apparent in
the figure).

\citet{rao03} and \citet{nestor02} suggest, from ground--based
multi-band imaging, that the low surface brightness emission detected
around the quasar (see Figure~\ref{fig:1127}$a$) could arise from a
foreground low surface brightness galaxy at the absorption
redshift. However, it is possible that the low surface brightness
signal is coming from the quasar host galaxy at $z_{em}=1.18$. The
background quasar is radio--loud and has strong X--ray emission. An
X-ray jet extends north east $236$~kpc from the quasar, projected
\citep{siemiginowska02,siemiginowska07}. There is a diffuse halo, both
detected in radio and X--ray, around the host quasar. It is not yet
clear whether the X-ray halo is real or a result of blurring from the
instrument PSF (Siemiginowska, A. 2007, private commutation). Thus, it
is possible that the low surface brightness detected by \citet{rao03}
may be from the background quasar. \citep{chun06} found a possible
underlying galaxy $0.6''$ from the quasar. Again, this may be a
foreground galaxy, or structure from the quasar host galaxy which is
commonly observed \cite[e.g.,][]{bahcall97}.

Galaxies G1 and G2 were targeted for this study.  These galaxies are
spatially separated by only $55.3$~kpc projected, and by a line of
sight velocity of $183$~\kms. G1 is an almost edge--on ($i=82^\circ$)
spiral that displays asymmetries on both sides of the galaxy. The
galaxy has $PA=21^\circ$.  It has absolute magnitude of $M_B=-20.40$,
which translates to $L_B=0.54L_B^\ast$.  The galaxy is at a projected
distance of $D=45.6$~kpc from the quasar line of sight. G2 is an
interesting galaxy; it has a major dust lane and a large bulge. Given
that we detected no emission lines, this galaxy could either be
classified as Sa or as an early--type S0 galaxy. G2 also has a tidal
disturbance along the major axis of the galaxy towards the north. The
galaxy has an inclination of $i=73^\circ$ and $PA=61^\circ$. It has
absolute magnitude of $M_B=-20.57$ which translates to
$L_B=0.63L_B^\ast$.  The galaxy is at a projected distance of
$D=81.0$~kpc from the quasar.

The rotation curve of G1, obtained from the {\OII}, $\rm{H}\alpha$,
and the [\NII] doublet, is presented in Figure~\ref{fig:1127}$b$.  The
maximum observed rotational velocity is $204$~{\kms}.  The rotation
curve of G2, obtained from the {\NaI} absorption doublet, is presented
in Figure~\ref{fig:1127}$c$.  The maximum observed rotational velocity
of G2 is $90$~{\kms}. Below both rotation curves, the {\MgII} and the
{\MgI} absorption profiles are shown on the velocity scale defined by
G1. The mean absorption redshift is offset by $-112$~{\kms} from the
systemic velocity of G1.  The {\MgII} absorption can be broken up into
two kinematic components. The first large saturated component has a
velocity spread of roughly 235~{\kms}. The second component,
blue--ward of the main component, contains a few weak clouds and has a
velocity spread of roughly 68~\kms. In the {\MgI} profile, one can
resolve the individual clouds that are saturated in {\MgII}. Almost
all of the {\MgI} gas is aligned with the saturated {\MgII}
component. Only a very weak {\MgI} cloud is detected in a second
kinematic component. The main component of the {\MgII} gas aligns with
the blue--ward wing of the G1 rotation curve, which is on the side
toward the quasar. The {\MgII} absorption velocities start at the
systemic velocity of G1 and is centered on the maximum galaxy rotation
velocity. Since the rotation curve of G2 is derived from {\NaI}
absorption lines, it is likely that the maximum rotation speed of the
galaxy extends to larger velocities then detected. The velocities of
G2 are consistent a portion of the large saturated component of the
{\MgII} and the small blue--ward clouds.

\subsection{Q2206-199 G1}

A WFPC--2/{\it HST\/} image of the Q$2206-199$ field is shown is
Figure~\ref{fig:2206}$a$.  The $z=1.0166$ galaxy, G1, was targeted for
this study. The galaxy was spectroscopically identified by
\citet{bergeron92}. The {\MgII} absorption was first reported by
\citet{sargent88}. The HIRES/Keck spectrum was originally presented by
\citet{prochaska97}.  The {\MgII} equivalent width is
$W_r(2796)=1.057$~{\AA}.

The spiral galaxy G1 is quite unusual in brightness and
morphology. The galaxy has an absolute magnitude of $M_B=-23.8$ which
translates to $L_B=8.0L_B^\ast$. Galaxies with super-$L^\ast$
luminosities are quite rare and represent only a few percent of the
galaxy population.  The galaxy spiral structure is tightly wound with
a large leading arm.  There appears to be numerous {\HII} regions. The
bulge is compact and offset from the isophotal center. Note that at
the galaxy redshift, the rest--frame mean wavelength of the F702W
filter is around $3440$~\AA, roughly rest-frame U--band. The galaxy
inclination is $i=57^\circ$ with a $PA=67^\circ$. The galaxy is at a
projected distance of $D=104.6$~kpc from the quasar.

The rotation curve, obtained from the {\OII} doublet, is presented in
Figure~\ref{fig:2206}$b$.  The maximum observed rotational velocity is
$\sim 40$~{\kms}. Given the low observed rotation velocities, the
galaxy is likely more face--on than the GIM2D model inclination
suggests. The galaxy morphology is asymmetric and unusual which makes
it difficult to determine the inclination.  Below the rotation curve,
the {\MgII} and the {\MgI} absorption profiles are shown on the same
velocity scale. The mean absorption redshift is offset by $73$~{\kms}
from the galaxy systemic velocity.  The {\MgII} absorption can be
broken up into two kinematic components. The first one has velocity
spread of roughly 150~{\kms} and is mostly saturated. The second
component is a single cloud with a velocity spread of roughly
28~{\kms} that is offset 198~{\kms} red--ward of the main component
velocity center. From the {\MgI} profile, one can resolve the
individual clouds that are saturated in {\MgII}. Most of the {\MgI}
gas is shifted $100$~{\kms} from the systemic velocity of the
galaxy. The red--ward arm the the rotation curve aligns with only a
small portion of velocity of the main {\MgII} component. However,
there is a large amount of {\MgII} gas that has greater velocities
than those of the galaxy. The bulk of the more neutral/high column
density gas, as indicated by {\MgI}, has velocities greater than that
of the observed galaxy rotational velocities.

\subsection{Summary I: Observational Kinematic Comparisons}

In our direct comparison of galaxy disk and halo gas kinematics,
traced by {\MgII} and {\MgI} absorption, we find the following: (1) in
all ten cases, the observed galaxy rotation velocities show
substantial overlap with the bulk of the absorption velocities.  (2)
in seven of ten cases, the {\MgII} and {\MgI} absorption resides to
one side of the galaxy systemic velocity. In the remaining cases,
(Q$0002+051$ G1, Q$0836+113$ G1, and Q$1127-145$ G3) absorption
resides on both sides of the galaxy systemic velocity.  Our findings
are similar to those of \citet{steidel02}, even though their sample
targeted highly inclined and edge--on disk galaxies with $PA\sim
0$. Here we have attempted to select galaxies with a range of
inclination and position angle with respect to the quasar line of
sight (see Table~\ref{tab:params}).

To see if there are differences in the {\MgII} absorption profiles as
a function of inclination, we have separated the galaxies into two
inclination bins with $i<60^\circ$ (five galaxies) and $i>60^\circ$
(six galaxies). The galaxies from the \citet{steidel02} sample are
included here.  For this comparison, multiple galaxies that can be
associated with a single absorption system (such as G1 and G2 in
Q$0450-132$ and $1127-145$) were removed since we cannot confidently
know whether one or both galaxies host the {\MgII} absorption.

\begin{figure}
\includegraphics[angle=0,scale=0.85]{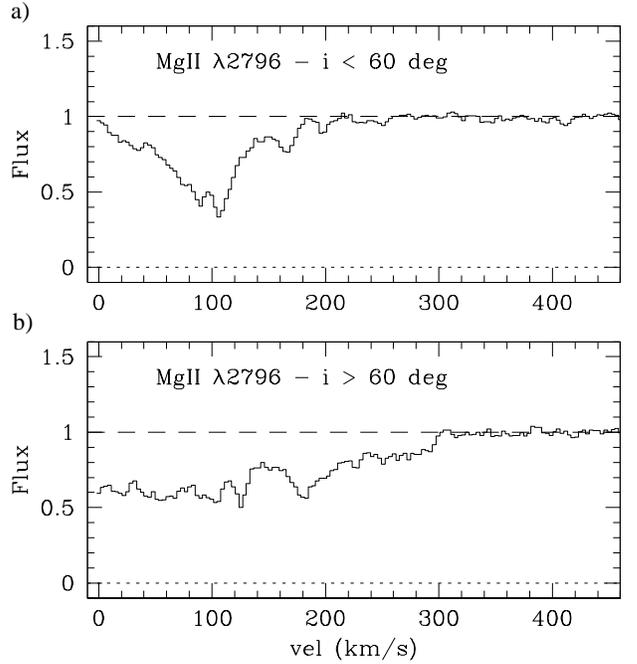}
\caption[angle=90]{The combined {\MgII} absorption spectra for
galaxies separated into two inclination bins. The spectra are plotted
versus absolute velocity difference from the galaxy systemic velocity.
The five galaxies of \citet{steidel02} are included. The two galaxy
pairs, G1 and G2 of Q$0450-132$ and Q$1127-145$, are excluded
here. --- (a) The summed spectra for five galaxies with
$i<60^\circ$. --- (b) The summed spectra for six galaxies with
$i>60^\circ$.}
\label{fig:ploti}
\end{figure}

In Figure~\ref{fig:ploti}, we show the co--added {\MgII} absorption
profiles. The data are plotted as an absolute velocity difference from
the galaxy systemic velocity.  As seen in Figure~\ref{fig:ploti}$a$,
the combined spectra of the galaxies with $i<60^\circ$ shows that
absorption resides between $0-180$~{\kms} with a peak in the optical
depth at $v\sim100$~{\kms}.  In Figure~\ref{fig:ploti}$b$, the
combined spectra show that galaxies with $i>60^\circ$ are associated
with {\MgII} absorption with smoothly varying optical depth spread
over $\sim300$~{\kms}.  A Kolmogorov-Smirnov (K-S) test reveals the
probability of the two optical depth distributions being drawn from
the same sample is $P_{\mbox{KS}}=0.000873$. This rules out the null
hypothesis of similar {\MgII} optical depth distributions as a
function of inclination at the 99.91\% confidence level ($3.3\sigma$).
The data are suggestive that {\MgII} absorption velocity spread may be
a function of galaxy orientation; galaxies with higher inclination
have {\MgII} absorption with larger velocity spreads and more evenly
distributed optical depths (on average at any given velocity).
However, since the number of galaxies per inclination bin is small, we
need to acquire a larger sample to see if the trend holds.

A trend with inclination might be expected if the absorbing gas
kinematics is well represented by a monolithic rotating halo. {\MgII}
absorption velocity spreads in four out of the five galaxies in the
\citet{steidel02} sample were shown to be consistent with a monolithic
rotating halo model that allowed for lagging rotation with increasing
height above the disk.  Given the apparent kinematic trend with
inclination, we investigate whether the lagging halo model
\citep{steidel02} can successfully predict the {\MgII} absorption
velocity spreads of our ten galaxies.


\section{Galaxy Kinematics and Halo--Disk Models}\label{halo}

We apply the simple halo model of \citet{steidel02} to our systems in
order to determine whether an extended disk--like rotating halo is
able to reproduce all or most of the observed {\MgII} absorption
velocity spread. The model is a co--rotating disk with velocity
decreasing as a function of scale height. 

The line of sight velocity, $v_{los}$, predicted by this disk
halo model is a function of the measurable quantities $D$, $i$, $PA$
(the angle between the galaxy major axis and the quasar line of
sight), and $v_{max}$, which is the maximum projected galaxy rotation
velocity,

\begin{eqnarray} 
\label{eq:kine}
v_{los}&=&\frac{-v_{max}}{\sqrt{1+\left(\displaystyle \frac{y}{p}\right)^2}}\mbox{~}
\exp \left\{-\frac{\left|y-y_{\circ}\right|}{h_{v}\tan i}\right\}\mbox{~~}\mbox{where}, \\
\nonumber\\[2.0ex]
y_{\circ}&=&\frac{D\sin PA}{\cos i} \mbox{~~~~~}\mbox{and}\mbox{~~~~~}p=D\cos PA\mbox{~},\nonumber
\end{eqnarray}



\noindent where the free parameter, $h_{v}$, is the lagging gas
velocity scale height and where $y$ is the projected line of sight
position above the disk plane. The parameter $y_{\circ}$ represents
the position at the projected mid--plane of the disk. The range of $y$
values is constrained by the model disk--halo thickness, $H_{\rm
eff}$, such that $y_{\circ}-H_{\rm eff}\tan i \leq y \leq
y_{\circ}+H_{\rm eff}\tan i$.  The distance along the line of sight
relative to the point were it intersects the projection of the disk
mid--plane is then $D_{los}=(y-y_{\circ})/\sin i$.  There are no
assumptions about the spatial density distribution of {\MgII}
absorbing gas, except that $H_{\rm eff}$ is the effective thickness of
the gas layer capable of giving rise to absorption.

In order to maximize the rotational velocity predicted by the model,
we assume $h_v=1000$~kpc, which effectively removes the lagging halo
velocity component (such that the exponential in
Equation~\ref{eq:kine} is roughly equal to unity).


In Figure~\ref{fig:toym}$a-h$, we show the {\MgII} absorption profiles
for each galaxy, where the shaded regions indicate detected
absorption. Below each absorption profile is the disk halo model
velocities as a function of $D_{los}$ derived for each galaxy (solid
line) using Equation~\ref{eq:kine} and parameters in
Table~\ref{tab:params}. Recall that, at $D_{los}=0$~kpc, the model
line of sight intersects the projected mid--plane of the galaxy.  The
dashed curves represent the disk halo model velocities derived from
the combination of the minimum and maximum uncertainties in the $PA$
and $i$. In some cases (see Figure~\ref{fig:toym}$b$) the values of
the $PA$ and $i$ are well determined such that the dashed curves lie
on the solid curves.  The model also predicts the line of sight
position, $D_{los}$, of the halo gas at each velocity, $v_{los}$.

The disk halo model is successful at predicting the observed
{\MgII} absorption velocity distribution when the solid (or dashed)
curves span the same velocity spread as that of the {\MgII} absorption
gas.  The model curves must occupy the full shaded region to be 100\%
successful.  If this is not the case, one can conclude that disk--like
halo rotation is not the {\it only\/} dynamic mechanism responsible
for the {\MgII} kinematics. In the following subsections we discuss
the disk model of the individual galaxies.
 

\subsection{Q0002+051 G1}\label{sec:0002m}

\begin{figure*}
\begin{center}
\includegraphics[angle=0, scale=1.00]{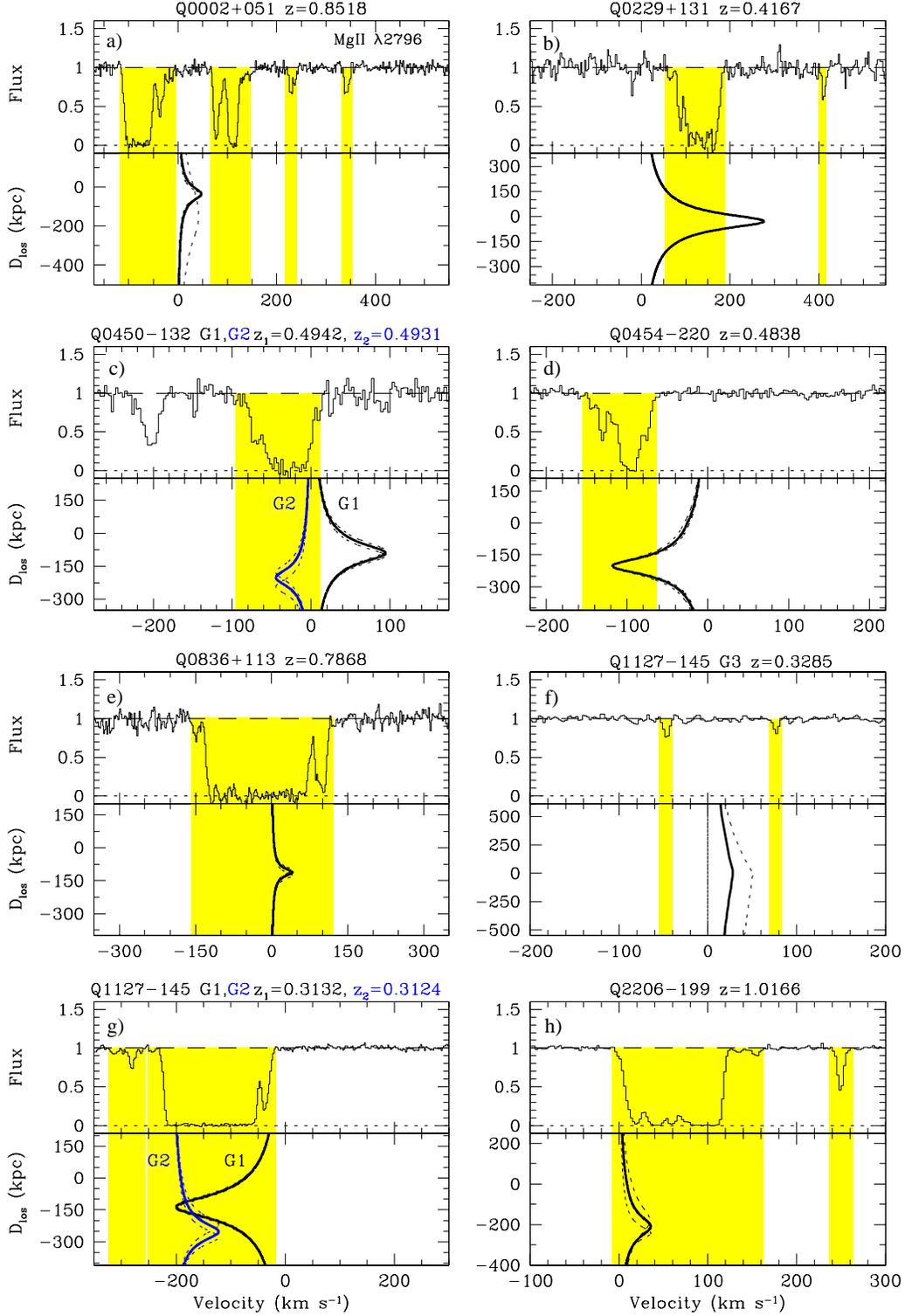}
\end{center}
\caption[angle=90]{The {\MgII} absorption profiles and the disk
model velocities as a function of $D_{los}$ (solid curve) are shown for
each galaxy in the top and bottom panels, respectively. The {\MgII}
absorption velocities are shaded in. The solid curve is computed using
Equation~\ref{eq:kine} and the values from Table~\ref{tab:params}. The dashed curves
are models computed for the maximum and minimum predicted model
velocities given the uncertainties of $i$ and $PA$. The disk model
is successful and reproducing the observed absorption velocities in the
solid curve overlaps with the entire shaded region. $D_{los}$ is
equal zero when the quasar line of sight intersects the projected
mid--plane of the galaxy. The panels are as follows; (a) Q$0002+051$
G1, G1 (b) Q$0229+131$ G1, (c) Q$0450-132$ G1 and G2, (d) Q$0454-220$
G1, (e) Q$0836+113$ G1, (f) Q$1127-145$ G3, (g) Q$1127-145$ G1 and G2,
and (h) Q$2206-199$ G1.}
\label{fig:toym}
\end{figure*}

The galaxy G1 exhibits a low level velocity shear.  Given the velocity
spread of the gas ($\sim 475$~\kms), it is impossible for the bulk of
the absorption gas to be consistent with the observed velocities of
G1. In Figure~\ref{fig:toym}$a$, we see that the galaxy disk
halo model is counter rotating with respect to the dominate saturated
{\MgII} component. There is no overlap between the predicted halo
model velocities with those of the {\MgII} and {\MgI} absorption.
Even if the galaxy had a highly significant velocity shear, the bulk
of the {\MgII} clouds would not be consistent in velocity space.
Given the number of high velocity components, it is unclear that this
absorption profile represents a gravitationally bound gaseous galactic
halo.

\begin{deluxetable}{lcllll}
\tabletypesize{\scriptsize}
\tablecaption{Galaxy Disk Model Input Values\label{tab:params}}
\tablecolumns{6}
\tablewidth{0pt}

\tablehead{
\colhead{ }&
\colhead{Galaxy} &
\colhead{$D$} &
\colhead{$v_{max}$} &
\colhead{$i$} &
\colhead{$PA$} \\
\colhead{QSO Field}&
\colhead{ID} &
\colhead{(kpc)} &
\colhead{(km/s)}&
\colhead{(deg.)} &
\colhead{(deg.)}
}
\startdata
Q$0002+051$ & G1 & $25.9\pm0.5$ & $-49$ & $38_{-31}^{+12}$& $43_{-6}^{+14}$\\
            &    &              &       &                 &               \\[-1.0ex]
Q$0229+131$ & G1 & $37.5\pm0.5$ & $-281$& $58_{-1}^{+2}$  & $22_{-2}^{+2}$\\
            &    &              &       &                 &               \\[-1.0ex]
Q$0450-132$ & G1 & $50.1\pm0.4$ & $-98$ & $66_{-2}^{+3}$  & $42_{-3}^{+2}$\\
            &    &              &       &                 &               \\[-1.0ex]
Q$0450-132$ & G2 & $62.7\pm0.7$ & $47$  & $75_{-2}^{+2}$  & $54_{-2}^{+2}$\\
            &    &              &       &                 &               \\[-1.0ex]
Q$0454-220$ & G1 & $107.9\pm0.8$& $138$ & $41_{-2}^{+1}$  & $76_{-2}^{+1}$\\
            &    &              &       &                 &               \\[-1.0ex]
Q$0836+113$ & G1 & $26.9\pm0.9$ & $-42$ & $78_{-1}^{+1}$  & $57_{-1}^{+1}$\\
            &    &              &       &                 &               \\[-1.0ex]
Q$1127-145$ & G1 & $45.6\pm0.3$ & $204$ & $82_{-0}^{+0}$  & $21_{-0}^{+0}$\\
            &    &              &       &                 &               \\[-1.0ex]
Q$1127-145$ & G2 & $81.0\pm0.3$ & $-90$ & $73_{-3}^{+1}$  & $61_{-2}^{+2}$\\
            &    &              &       &                 &               \\[-1.0ex]
Q$1127-145$ & G3 & $91.4\pm0.2$ & $-80$ & $1_{-1}^{+3}$   & $69_{-19}^{+34}$\\
            &    &              &       &                 &               \\[-1.0ex]
Q$2206-199$ & G1 & $104.6\pm1.4$& $-40$ & $57_{-14}^{+5}$ & $67_{-6}^{+7}$\\
            &    &              &       &                 &               
\enddata
\end{deluxetable}

\subsection{Q0229+131 G1}\label{sec:0229m}

The galaxy G1 has an asymmetric rotation curve with the largest
rotation velocity observed in the direction of the quasar line of
sight. In Figure~\ref{fig:toym}$b$, we see that the disk halo
model velocities are consistent with the bulk of the {\MgII}
absorption. Thus, extended disk--like halo rotation could be invoked
to explain most of the observed halo gas velocities. However, there
remains a small single {\MgII} halo cloud $\sim 400$~{\kms} from
systemic that that cannot be explained by halo rotation alone. This
suggests other dynamic processes give rise to some of the {\MgII}
absorption.

\subsection{Q0450-132 G1, G2}\label{sec:0450m}

Galaxies G1 and G2 are potentially interacting galaxies, as evident
from their morphologies and strong emission lines.  It is possible
that these interactions are an effective mechanism in producing
extended {\MgII} absorption in the halo
\citep{bowen95,kacprzak07,rubin09}.  In Figure~\ref{fig:toym}$c$, we
plot the disk halo models for both galaxies. G1, the galaxy closest to
the quasar, has model halo gas kinematics that are counter--rotating
with respect to the {\MgII} absorption. G2, on the other hand, has
modeled halo velocities that are consistent with those of the {\MgII}
absorption. The halo model of G2 is also consistent with the bulk of
the {\MgI}. Given that G1 was observed along the major axis and G2 was
not, if we assume that G2 had comparable rotation speeds as G1, the
halo model velocities would overlap with most of the absorption
velocities.  We will discuss the difficulties of disentangling these
multiple galaxy systems in \S~\ref{sec:double}.

\subsection{Q0454-220 G1}\label{sec:0454m}

Galaxy G1 has a symmetric rotation curve that completely flattens out
at the maximum velocity of $138$~{\kms}. In Figure~\ref{fig:toym}$d$,
we see that the disk halo model has velocities consistent with
the bulk of the {\MgII} absorption velocities. They are also
consistent with the {\MgI} absorption velocities. However, there is an
inconsistency of $\sim40$~{\kms} between the halo model and {\MgII}
absorption velocities. Thus, the halo model is unable to reproduce the
total observed spread of {\MgII} absorption velocities.

\subsection{Q0836+113 G1}\label{sec:0836m}

The galaxy G1 exhibits minimal rotation; the velocities are more
indicative of a global shear. The galaxy systemic velocity is centered
roughly in the middle of the {\MgII} and the bulk of the {\MgI}
absorption profiles. In Figure~\ref{fig:toym}$e$, we see that the halo
model also shows little rotation, roughly $50$~{\kms}. If the model
was a true representation of the halo, then more than 50\% of the
absorbing gas has velocities inconsistent with disk rotation that are
larger than the model velocities. Even if G1 has a more significant
velocity shear, the model would still not be able to explain the gas
blue--ward of the galaxy systemic velocity. Here, the models fails to
predict the bulk of the absorption velocities.

\subsection{Q1127-145 G3}\label{sec:1127mG3}

The galaxy G3 appears face--on, however it exhibits a maximum rotation
of $80$~{\kms}. The model inclination of $1^\circ$ is likely incorrect
given the observed rotation velocities.  In Figure~\ref{fig:toym}$f$,
the disk halo model for G3 exhibits little line of sight velocity
($\sim20-50$~\kms) given its orientation with respect to the quasar
line of sight. Given the model parameters listed in
Table~\ref{tab:params}, we varied the galaxy inclination, such that
$0\leq i \leq 45^\circ$, in an attempt to reproduce the observed
absorption velocities. Even with $i=45^\circ$, the halo model fails to
reproduce the observed {\MgII} absorption velocities. For the disk
halo scenario, the halo gas is expected to reside to one side of the
rotation curve. The nature of this {\MgII} absorption profile is
interesting; two weak clouds separated by $125$~{\kms}. It is likely
that these clouds could arise in either a patchy diffuse halo or the
line of sight is intercepting small scale structure near the galaxy
halo. In any case, the model is unable to reproduce the observed
{\MgII} absorption velocities.

\subsection{Q1127-145 G1,G2}\label{sec:1127m}

Galaxies G1 and G2 are potentially interacting galaxies, as evident
from their morphologies. These galaxies appear to be in a small group
including G4. In Figure~\ref{fig:toym}$g$, we plot the disk halo
models for G1 and G2.  The model for G1 has velocities that are
consistent with up to $200$~\kms spread of the saturated component of
the {\MgII} and all of the {\MgI} absorption. The disk halo
model of G2 is counter--rotating with respect to G1 as viewed from the
quasar line of sight. The model velocities are consistent with the
remaining absorption of the saturated component which was not covered
by G1. The model velocities of both G1 and G2 overlap
$\sim75$~{\kms}. The second weaker kinematic component,
$\sim300$~{\kms} blue--ward of systemic velocity of G1, cannot be
explained given the predicted halo velocities.  It is possible that
the saturated component of the {\MgII} absorption could arise from
either G1, G2, or both. The weaker component may arise from tidal
debris stirred up by the apparent interactions. It is also possible
that some of the {\MgII} absorption is associated with G4.  It is
likely that the saturated component is associated with a nearby galaxy
since it is a DLA and many other low ions have also been detected. We
will discuss the difficulties of disentangling these multiple galaxy
systems in \S~\ref{sec:double}. In any case, the model is mostly
successful, except that it does not reproduce all of the {\MgII}
absorption velocities.

\subsection{Q2206-199 G1}\label{sec:2206m}

The G1 galaxy is the largest in our sample, and the second furthest
away from the quasar line of sight. In Figure~\ref{fig:toym}$h$, the
disk halo model shown has very little line of sight velocity
($\sim40$~\kms). It is clear that the model of this moderately
inclined galaxy does not reproduce the observed absorption
velocities. In fact, the bulk of the {\MgII} and {\MgI} has velocities
$\sim 50$~{\kms} greater than the galaxies maximum observed rotation
velocity.  This is peculiar, since the dominant saturated component is
commonly expected to be associated with the galaxy disk, yet the
kinematics here suggest otherwise.  This galaxy--absorber pair is a
another example demonstrating that disk--like halo rotation cannot be
the only mechanism driving the kinematics of halo gas.

\subsection{{\MgII} Absorption from Galaxy Pairs/Groups}\label{sec:double}

Since {\MgII} absorbers were first associated with galaxy halos, it
has been common practice to associate one galaxy with an absorption
system at a given redshift \citep[e.g.,][]{bb91,sdp94,csv96,gb97}. It
is now becoming evident that {\MgII} absorption also arises in small
groups of galaxies \citep{bowen95,cwc99} and even clusters
\citep{lopez08}. In this paper, we present two such examples:
Q$0450-132$ G1, G2 and Q$1127-145$ G1, G2, G4. The pair of galaxies in
Q$0450-132$ are close to each other in both projected distance and
velocity. Both galaxies have morphological evidence (one sided tidal
tails) that is classically associated with interacting/merging
galaxies. The Q$1127-145$ field contains three galaxies at similar
redshift (we have recently spectroscopically identified two
additional galaxies with similar redshifts within $D=250$~kpc from the
quasar [Kacprzak et al., in prep]). The two galaxies studied here, G1
and G2, have evidence of morphological perturbations and extended
tidal material.

The Q$0450-132$ and Q$1127-145$ galaxy pairs, have observed rotation
velocities that overlap with those of the {\MgII} absorption. Given
these two fields, it is clear that it can be difficult at times to
associate one particular galaxy with an absorption system. One
alternative interpretation is that the material responsible for the
absorption is tidal debris originating from both galaxies due to past
mergers and harassments (similar to the Magellanic stream). Given that
the galaxies are in close proximity (projected), it is also possible
that these galaxies share a common gas structure that gives rise to
the absorption.

\subsection{Summary II: Disk Halo Model}


In an effort to reproduce the {\MgII} absorption velocities, we used a disk halo model to compute the expected absorption velocities.  In only one case, Q$0229+131$ G1, we were able to reproduce the full spread of the {\MgII} absorption velocities in a disk halo model.  In four other cases, Q$0454$--$220$, Q$0450$--$132$ (G2), and Q$1127$--$145$ (G1 and G2), the velocity region with the strongest absorption is consistent with the model.  However, the halo model of the galaxy G2 toward Q$0450$--$132$ does not account for roughly 35\% of the absorption.   In the cases of both Q$0229+131$ and Q$1127-145$, the models  cannot account for the unsaturated small cloud structures at higher velocities relative to systemic.   For each case, the models do reproduce some of the absorption velocities, however, the disk rotating halo model is insufficient to account for the full observed {\MgII} absorption velocity range.

We emphasize that the disk halo model applied to the data is an extreme version of the spatial--kinematic relationship in that (1) all
the gas is assumed to rotate at the maximum observed velocity of the galaxy, and (2) the scale height of the models ($h_v = 1000$ kpc) is highly unrealistic.   These extreme conditions were required in order to obtain the greatest degree of agreement between the model and the data.  If these conditions are relaxed, the level of agreement we found is diminished substantially.   None the less, even under these extreme and unrealistic model parameters, the disk halo model provides insight into the degree at which rotation kinematics can account for limited regions of the absorption velocity spread.   

What we learn from the disk halo model is that it is reasonable to suggest that additional dynamical processes (such as infall, outflow, supernovae winds, mergers, etc.)  and/or additional satellites or unidentified galaxies giving rise to some of the {\MgII} absorption contribute to the observed velocity spreads. The possibility of unidentified galaxies is difficult to quantify, for the assigning of a given galaxy to a given absorption system is by its very nature not 100\% certain.  It would, however, require the galaxies we have assigned to not be a significant contributor to the absorption and/or the unidentified galaxy to have very extended absorbing gas.  The former statement is not strongly supported by a body of previous studies \citep[e.g.,][]{bb91,lebrun93,sdp94,gb97,s97,cwc-china,tripp-china,zibetti06,chen08,tinker08} The later statement is based upon the detected galaxies in the quasar fields.  The statement would not apply to putative galaxies below our detection limit that are in close projection to the quasar.

\section{$\Lambda$CDM Cosmological Galaxy Simulations}\label{sec:sim}

To further understand the halo gas dynamics and the mechanisms driving
the observed gas velocity spread, we investigate high resolution
cosmological simulation of galaxy formation, which include the
dynamical processes of infall, outflow, supernovae winds, mergers,
etc.  Simulations provide the only theoretical means to fully
incorporate these dynamical processes in a cosmological setting. We
use the method of quasar absorption lines through the simulations to
``observe'' the {\MgII} absorption kinematics. Here we analyze a
single $z=0.9$ simulated galaxy in detail to study the possible
structures and mechanisms that give rise to the observed {\MgII}
halo gas kinematics. By comparing halo gas kinematics in the
simulations to the spatial and dynamic processes of the simulated
galactic environments, we can gain further insights into the observed
{\MgII} absorption properties.


\subsection{Description of The Simulations}

The $\Lambda$CDM cosmological simulations are performed using the
Eulerian Gasdynamics plus N--body Adaptive Refinement Tree (ART) code
\citep{kravtsov99,kravtsov03}. Physical processes implemented in the
code include star formation and stellar feedback, metal enrichment
from type II and Ia supernovae, self--consistent advection of metals,
and metallicity--dependent cooling and photoionization heating due to
a cosmological ultraviolet background \citep{haardt96}. The code does
not include radiative transfer, magnetic fields or Kelvin-Helmholtz
instabilities. The cosmological model has $\Omega_m=0.3,$
$\Omega_\Lambda = 0.7$, $h=0.7$, and $\sigma_8=0.9$.

These simulations have high star formation efficiency at the
resolution scale. This assumption is motivated by studies of star
formation in simulations of isolated disks with a similar resolution
\citep{tasker06}. The gas consumption time-scale is $\sim10^7$~yr for
the star-forming cells, however, only a small fraction ($\sim$1\%) of
the gas in the disk is forming stars in a typical time step of
$\sim$40 Myr at z$\sim$1. As a result, the disk-averaged star
formation efficiency is low: the gas consumption time-scale is
$\sim10^9$~yr, consistent with observations \citep{kennicutt98}.

The computational region is a 10 $h^{-1}$ Mpc co--moving box. We apply
a zooming technique \citep{klypin01} to select a Lagrangian volume of
three viral radii centered in a MW--size halo at redshift $z=0$. The
volume is then re--simulated from $z=50$ with higher resolution and
hydrodynamics. The high-resolution region has a radius of about
1.5~$h^{-1}$ co-moving Mpc, and has about $4 \times 10^5$ dark matter
particles with $5 \times 10^6$M$_{\odot}$ per particle. The volume is
resolved with about $1.3 \times 10^7$ hydrodynamic cells with
different levels of resolution. 
 
The combined effect of stellar winds and supernova explosions at the
resolved scale \citep{ceverino09} prevent the over--cooling problem of
galaxy formation at high redshifts \citep{white91} and reduce the
angular momentum problem found in early simulations
\citep{navarro00}. This results in models with flat rotation curves
consistent with observations \citep{ceverino09}, and is achieved
without typical ad--hoc assumptions about the physics at
sub-resolution scales. These high resolution simulations allow us to
resolve the regime in which stellar feedback overcomes the radiative
cooling. By resolving this regime, simulations naturally produce
galactic scale outflows in star-forming galaxies (Ceverino \& Klypin
2010, in preparation) and galaxy formation proceeds in a more
realistic, although violent way, through a combination of cold flows
accretion, mergers, and galaxy outflows.

\subsection{Simulated Spectra}

To study the {\MgII} absorption arising in the gas halos within the simulations, we employ the following methods.  For a given gas cell probed by a line of sight through the simulation box, the total hydrogen density, temperature, and metallicity is used to obtain the {\MgII} ionization fraction assuming photoionization conditions.  Post simulation, we use Cloudy \citep[V96b4,][]{ferland-cloudy} with the \citet{haardt96} UV background spectrum at the appropriate redshift.  The line of sight the redshift, {\MgII} column density, and Doppler $b$ parameter (assuming thermal broadening) are computed for each cell. 

Absorption spectra with the instrumental and noise characteristics of
the HIRES spectrograph are generated assuming each cell gives rise to
a Voigt profile at its line of sight redshift.  We give each spectrum
a signal--to--noise ratio of 50 per pixel, which corresponds to a
limiting equivalent width detection of 0.005~{\AA} for unresolved
lines.  The spectrum for each sightline is then objectively analyzed
for detectable absorption above the equivalent width threshold of
0.02~{\AA}, which corresponds to $\log N({\MgII}) = 11.7$~{\cmsq} for
$b=5$~{\kms}. The mean optical depths (mean redshifts), rest--frame
equivalent widths and velocity widths, and other quantities are then
measured \citep[see][]{cv01}.  The velocity zero point of the
simulated absorption lines is set to the line of sight velocity of the
simulated galaxy (center of mass of the stars).

To examine the 3D spatial and kinematic properties of gas giving rise
to {\MgII} absorption, we identify {\MgII} ``absorbing gas cells''
along each sightline as those which contribute to detected absorption
in the simulated spectra; they are defined as cells that align within
the range of line of sight velocities of the absorption.  We account
for multiple kinematics subsystems \citep{cv01}, regions of absorption
separated by continuum.

\begin{figure*}
\begin{center}
\includegraphics[angle=0,scale=0.63]{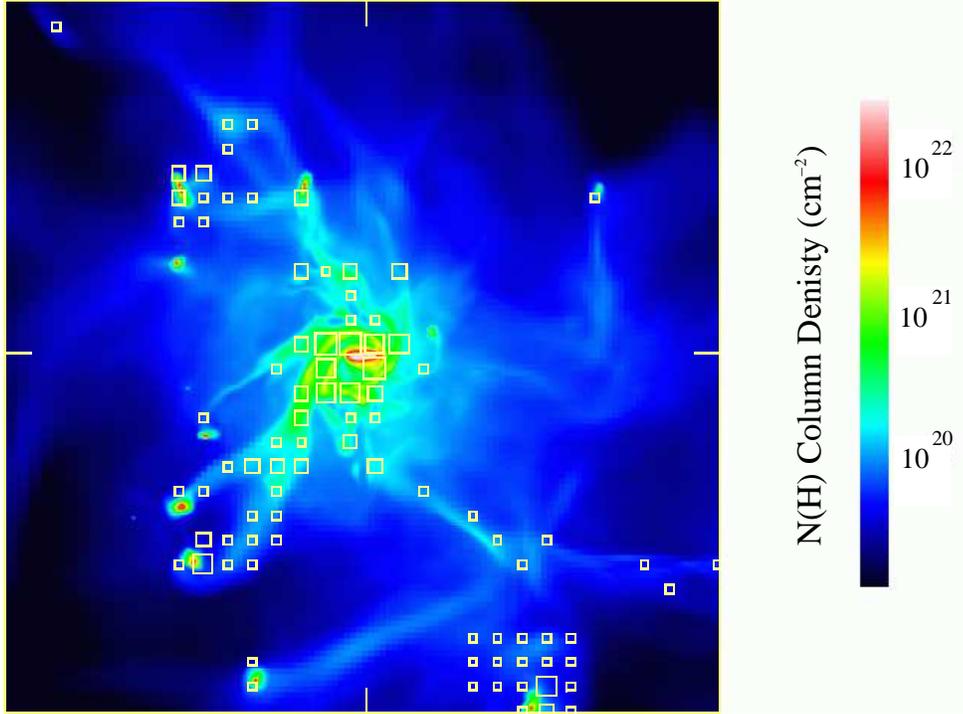}
\caption{The integrated total hydrogen column density, N(H), over a
220~kpc cube is shown for an $z=0.923$ simulated galaxy viewed edge
on. The direction of the simulated quasar lines of sight are
perpendicular to the plane of the image. Squares of increasing size
are plotted where {\MgII} absorption was detected along the line of
sight in the simulated quasar spectra. We apply an equivalent width
detection limit of $W_r(2796) \geq0.02$~{\AA}. The four square sizes
indicate, in increasing order, {\MgII} absorption equivalent width
bins of; $0.02 \leq W_r(2796) \leq 0.3$, $0.3 < W_r(2796) \leq 0.6$,
$0.6 < W_r(2796) \leq 1.0$, and $1.0 < W_r(2796) \leq 3.0$,
respectively.}
\label{fig:eonH}
\end{center}
\end{figure*}

\begin{figure*}
\begin{center}
\includegraphics[angle=0,scale=0.63]{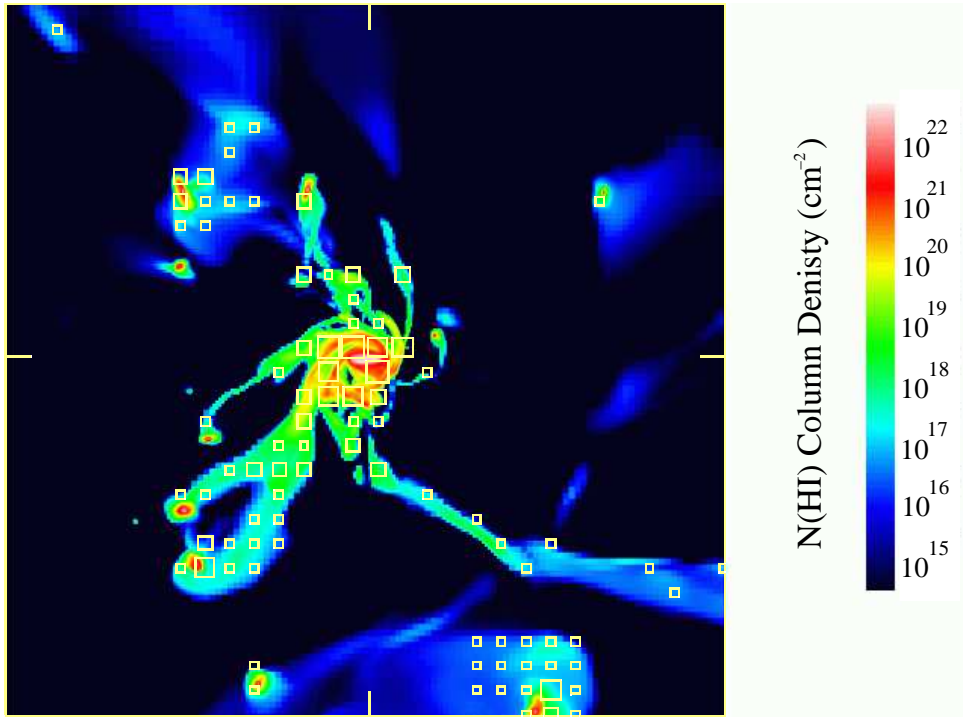}
\caption{Same as Figure~\ref{fig:eonH}, except the integrated neutral
{\HI} column density, N(\HI), is shown. The
direction of the simulated quasar lines of sight are perpendicular to
the plane of the image. Note that small DLA regions, having
$\hbox{N(\HI)}\geq 10^{20.3}$~{\cmsq}, are seen beyond $\sim85$~kpc.}
\label{fig:eonHI}
\end{center}
\end{figure*}
\begin{figure*}
\begin{center}
\includegraphics[angle=0,scale=0.63]{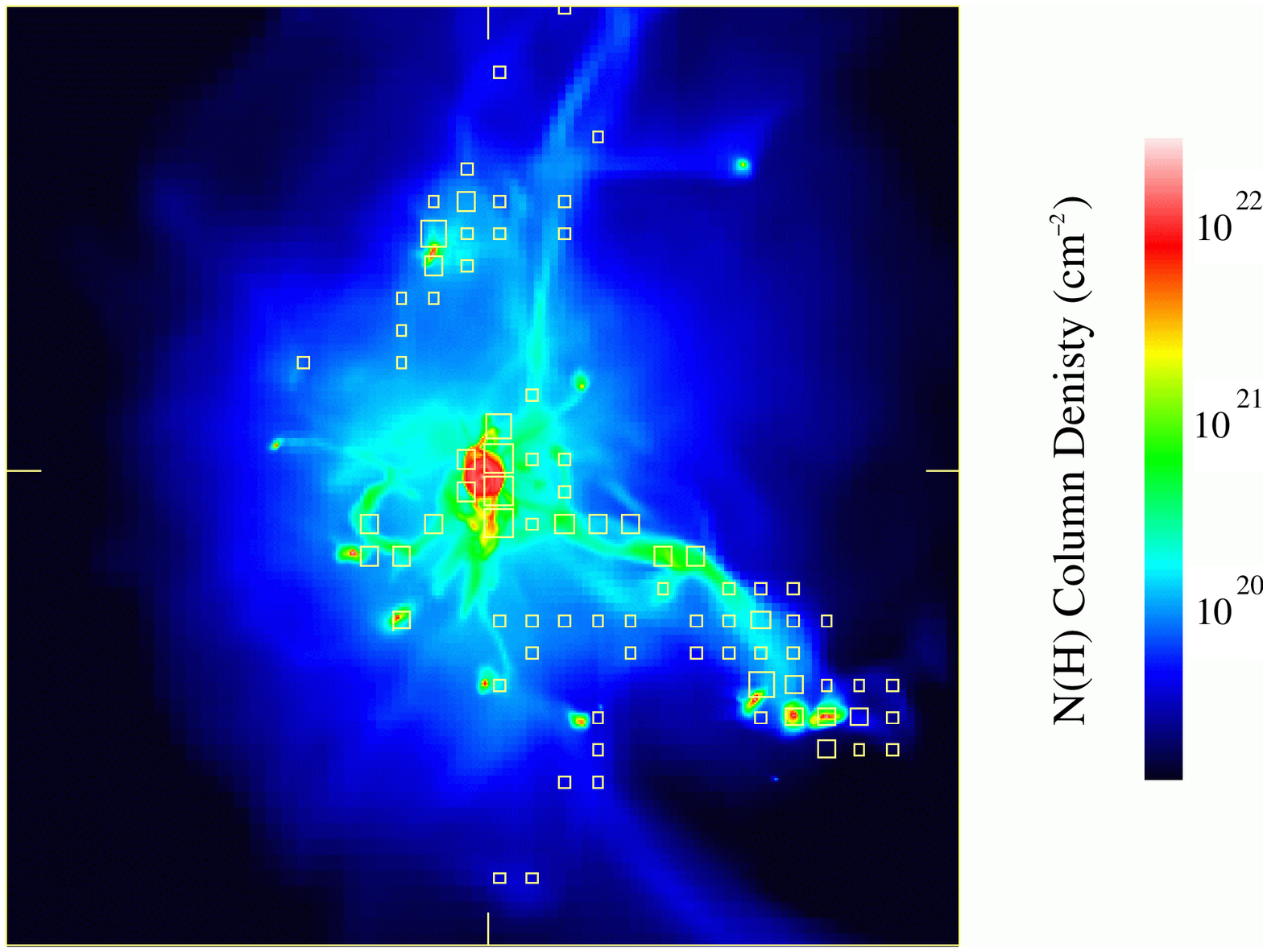}
\caption{Same as Figure~\ref{fig:eonH} for the same galaxy, except
viewed face on.  The direction of the simulated quasar lines of sight
are perpendicular to the plane of the image.}
\label{fig:fonH}
\end{center}
\end{figure*}

\begin{figure*}
\begin{center}
\includegraphics[angle=0,scale=0.63]{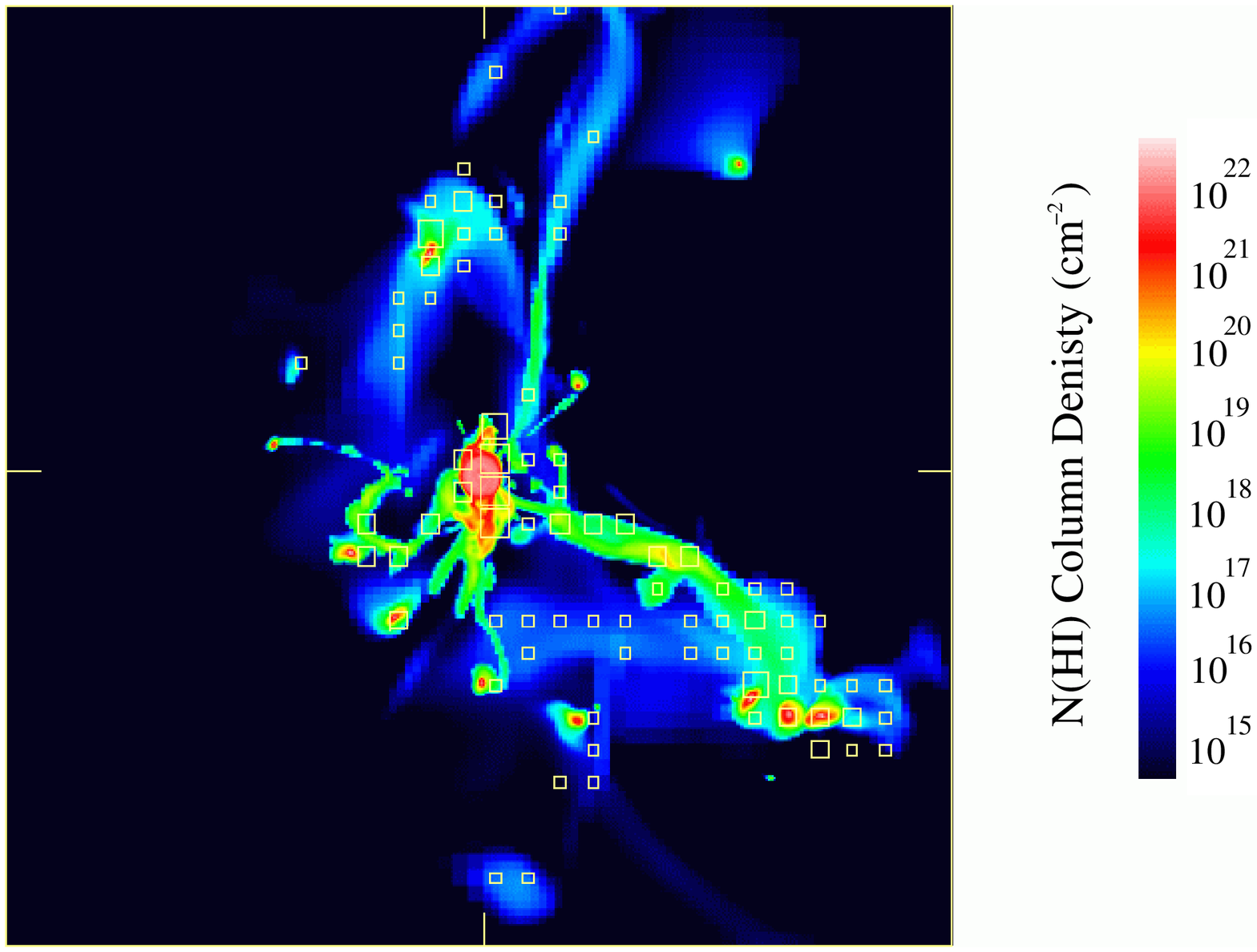}
\caption{Same as Figure~\ref{fig:eonHI} for the same galaxy, except
viewed face on. The direction of the simulated quasar lines of sight
are perpendicular to the plane of the image.}
\label{fig:fonHI}
\end{center}
\end{figure*}
\begin{figure*}
\begin{center}
\includegraphics[angle=0,scale=0.60]{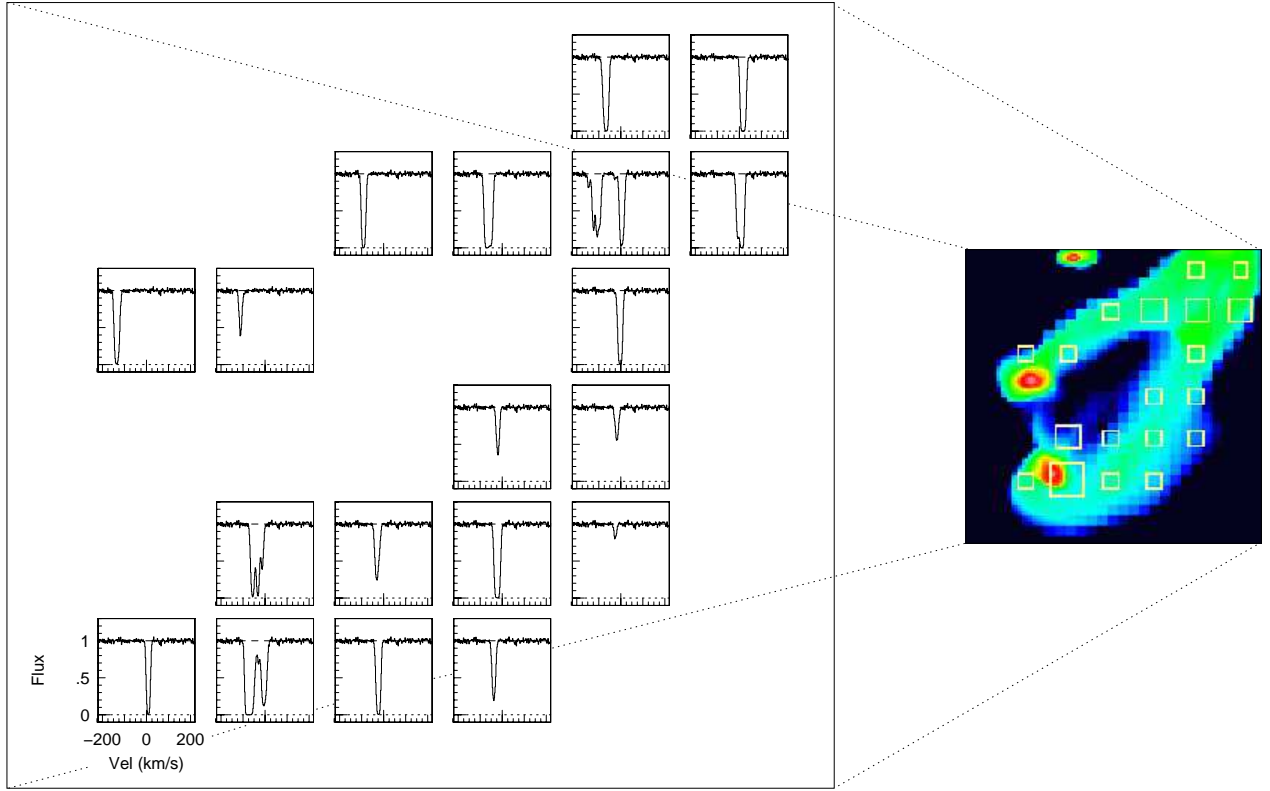}
\caption{A $45\times45$~kpc region extracted from the lower left
quadrant of Figure~\ref{fig:eonHI}. The region shows two satellite
galaxies and their tidal streams. The velocity zero point of the
absorption profiles is the galaxy systemic velocity. We enforced a
detection sensitivity limit of $W_r(2796)\geq0.02$~{\AA}. The 
majority of the {\MgII} absorption arising in this tidal stream is radially infalling
towards the galaxy.}
\label{fig:stream}
\end{center}
\end{figure*}

\begin{figure*}
\begin{center}
\includegraphics[angle=0,scale=0.60]{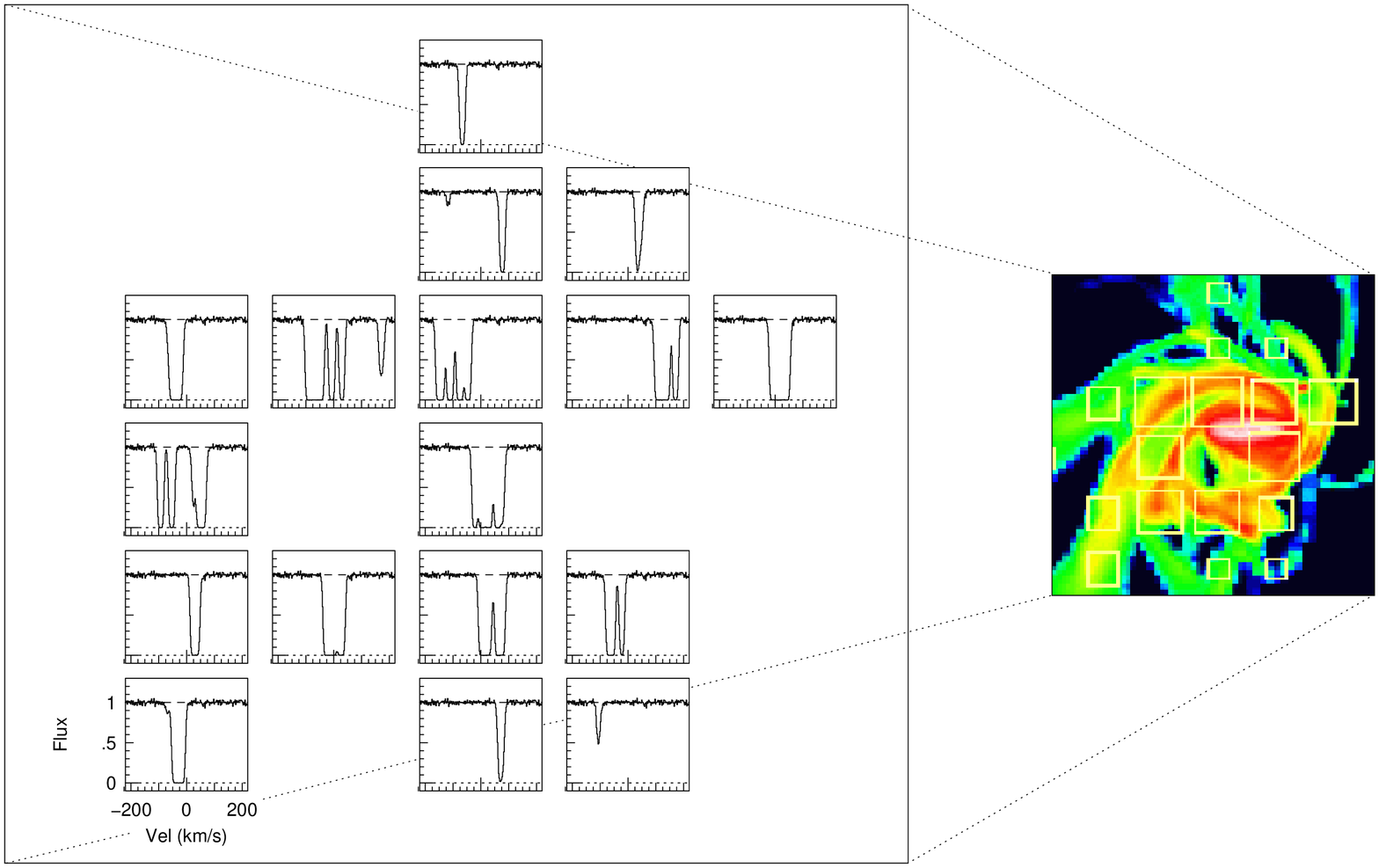}
\caption{A $37.5\times45$~kpc inner central region of the edge--on
  galaxy seen in Figure~\ref{fig:eonHI}.  The velocity zero point of
  the absorption profiles is the galaxy systemic velocity. We enforced
  a detection sensitivity limit of $W_r(2796)\geq0.02$~{\AA}. In the
  inner $\sim15$~kpc of the galaxy center we find some gas outflowing
  at $\sim200$~{\kms}. The {\MgII} absorption profiles produced by
  these outflows are similar to those seen for Q$0002+051$ G1
  (Figure~\ref{fig:0002_0229}b) and Q$0836+113$ G1
  (Figure~\ref{fig:0836_1127}b).}
\label{fig:center}
\end{center}
\end{figure*}

\subsection{Discussion of Simulated Galaxy Observations}

We focus on a single typical galaxy at $z=0.923$.  The galaxy star
formation rate is $SFR=3.5$~M$_{\odot}$~yr$^{-1}$.  The galaxy has a
maximum rotation velocity of 180~{\kms} when observed edge on. Based
upon the the Tully--Fisher relation, we derive a luminosity of
$L_B=0.4L_B^\ast$. The average luminosity for our sample is
$\left<L_B\right>=0.58L_B^\ast$ excluding the $8L_B^\ast$ galaxy.
Thus, this simulated galaxy is well representative of our
observational sample.

The simulated galaxy is probed with with a square grid of sightlines
at intervals of 7.5~kpc that span $-110$ kpc to $+110$ kpc on the
``sky'' and $14$~Mpc along the line of sight\footnote{For this and all
subsequent discussions, spatial quantities are quoted as proper
lengths}.  There are 900 total sightlines. The highest resolution of
the adaptive mesh at $z=0.923$ is $225$~pc. The gas contributing to
detectable {\MgII} absorption is found in a range of cell resolutions
from $225-1815$~pc with the majority of the gas arising in cells of
resolution 905~pc. We examine simulated quasar lines of sight for this
galaxy with three different inclinations, $i=5^{\circ}$ (face--on),
$45^{\circ}$, and $85^{\circ}$ (edge--on).

In Figure~\ref{fig:eonH}, we show the integrated total hydrogen column
density, N(H), over a $220$~kpc cube for the edge-on view of the
galaxy. There is no absorption detected along the line of sight
outside of this cube. The galaxy is clearly not in isolation.  In the
image, filaments and tidal stream material can be seen. Several low
mass satellites galaxies are also in the process of interacting with
the main galaxy.  We have superimposed squares over the sightlines
where {\MgII} absorption was detected in simulated quasar
spectra. Increasing square sizes indicate the absorption strength in
four bins: $0.02 \leq W_r(2796) \leq 0.3$, $0.3 < W_r(2796) \leq 0.6$,
$0.6 < W_r(2796) \leq 1.0$, and $1.0 < W_r(2796) \leq 3.0$. Out of the
900 lines of sight, {\MgII} absorption was detected in 87 for the
edge--on case.  Note that the absorption is not distributed
ubiquitously on the sky around the galaxy, but traces the various
structures around the galaxy. The covering fraction of the {\MgII} gas
is low ($\sim10$\%).

In Figure~\ref{fig:eonHI}, we show the integrated neutral hydrogen
column density, N(\HI), over the same cube.  The N(\HI) range from
roughly $10^{15}$ to $10^{22}$~{\cmsq}. The structures that are
associated with absorption stand out more in {\HI}.  We see as
expected \citep{weakI} that {\MgII} absorption is detected in regions
with N(\HI)$\gtrsim 10^{16.5}$~{\cmsq}.  Regions in which N(\HI)$ >
10^{16.5}$~{\cmsq} that fall between the line of sight grid sampling
are also expected to produced {\MgII} absorption. Note that as one
goes to lower N(\HI), the gas covering fraction increases.

In Figures~\ref{fig:fonH}--\ref{fig:fonHI}, we present N(H) and
N(\HI), respectively, for the face--on view of the galaxy. Out of the
900 lines of sight, {\MgII} absorption was detected in 96.  The
covering fraction remains roughly 10\%. Again, the absorption
primarily arises in streams and filaments.

We do not show the 45 degree view of the galaxy here. Out of the 900
lines of sight, {\MgII} absorption was detected in 124, and the
covering fraction is roughly $20$\%.

Overall, there are many complicated structures that reside within what
we classically call a halo. Note that we find pockets of DLA {\HI}
column densities [N(\HI)$>10^{20.3}$~{\cmsq}] out to
$\sim100$~kpc. Although, the covering fractions of these dense regions
are low, their presence suggests that it is possible to observe DLA
absorption at high impact parameters.  These large impact parameter
DLA systems arise from the inner regions of satellite galaxies.  These
satellites have a luminosity range $0.003L^\ast \leq L_B \leq
0.006L^\ast$. Assuming standard K-corrections, their apparent
magnitude in the {\it HST} F702W filter range between $ 28 \lesssim
m_{F702W}\lesssim 29$. These DLA producing satellites are well below
the typical detectability of WFPC--2/{\it HST} and ACS/{\it HST}.
Also note that the satellite in the upper right corner of
Figure~\ref{fig:fonHI} has an {\HI} morphology similar to galaxies
seen on their first pass through a cluster, where gas is being
stripped \citep[e.g.,][]{chung07}.

Focusing on the two small satellite galaxies and tidal stream in a
$45\times45$~kpc region of the lower left quadrant of
Figure~\ref{fig:eonHI}, we present an example of simulated {\MgII}
$\lambda 2796$ absorption spectra in Figure~\ref{fig:stream}.  The
velocity zero point of the simulated spectra is set to the galaxy
systemic velocity.  The absorption profiles are quite similar to those
detected in {\MgII} surveys. The simulated profiles have velocity
spreads of $\sim25-100$~{\kms}, some comprising multiple kinematic
components and complex structures.  The {\MgII} absorption spans
across the galaxy systemic velocity or resides blueward of the galaxy
systemic. Absorption arising near the galaxy systemic velocity or
entirely to one side is consistent with the majority of our
observational data.

In Figure~\ref{fig:center}, we show the {\MgII} $\lambda 2796$
absorption spectra over the $37.5\times45$~kpc central region of the
edge-on galaxy shown in Figure~\ref{fig:eonHI}. Individual absorption
profile velocities, in the inner regions of the galaxy, show a variety
of complex kinematics and optical depths.  In our observational sample
of galaxies, we have no galaxy--absorber systems that have impact
parameters less than $25$~kpc.  In the simulations, the inner
$\sim15$~kpc contains some $\sim200$~{\kms} outflowing gas.  The
outflows are not strong, but their signatures are reflected by the
complex kinematics of the absorption profiles in the inner
regions. The reason why there is a line of sight near the center of
the galaxy that does not produce {\MgII} absorption is because it is
dominated by $10^7$~K gas. The profiles reduce in kinematic complexity
rapidly as impact parameter increases.

To the degree that the simulations reflect the reality of the gaseous environments around galaxies, we find that {\MgII} absorption arises in many types of structures (metal enriched filaments, minor satellites, tidal streams, and within the region of the galaxy itself).  As inferred from our simulated absorption line survey, and guided by analysis of the 3D simulations (examples provided by Figures~\ref{fig:eonH}--\ref{fig:center}), the simulations are not suggestive of {\MgII} absorption arising from spherical or disk--like halos with uniform gas covering fraction. We revisit the spatial and kinematic distribution of the absorbing gas in \S~\ref{sec:gas}.

\begin{figure*}
\begin{center}
\includegraphics[angle=0,scale=0.45]{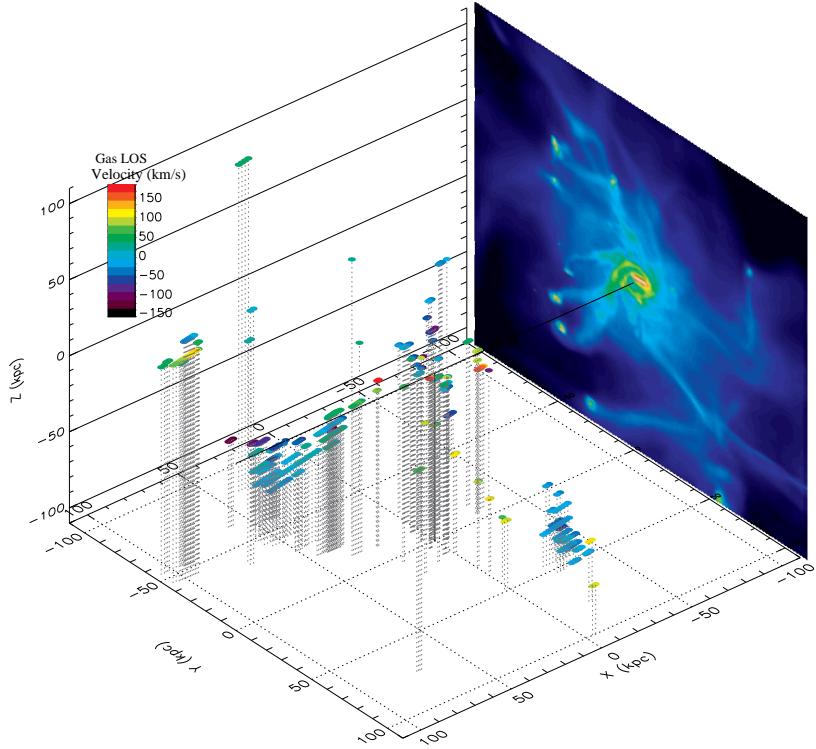}
\caption[angle=90]{The spatial distribution of N(\MgII)$<11.5$~{\cmsq}
  gas contributing to the {\MgII} absorption along the lines of sight.
  The edge--on galaxy is located at the origin; the black horizontal
  line represents a line of sight passing through the galaxy center.
  The observer is looking along the positive $x$ direction.  All
  simulated lines of sight are parallel to the $x$--axis. The
  absorbing gas is color coded as a function of line of sight velocity
  relative to the galaxy systemic velocity, as coded in the legend.
  Red absorbing gas is moving away from the observer; blue is moving
  toward.}
\label{fig:plotVabs}
\end{center}
\end{figure*}

\begin{figure*}
\begin{center}
\includegraphics[angle=0,scale=0.45]{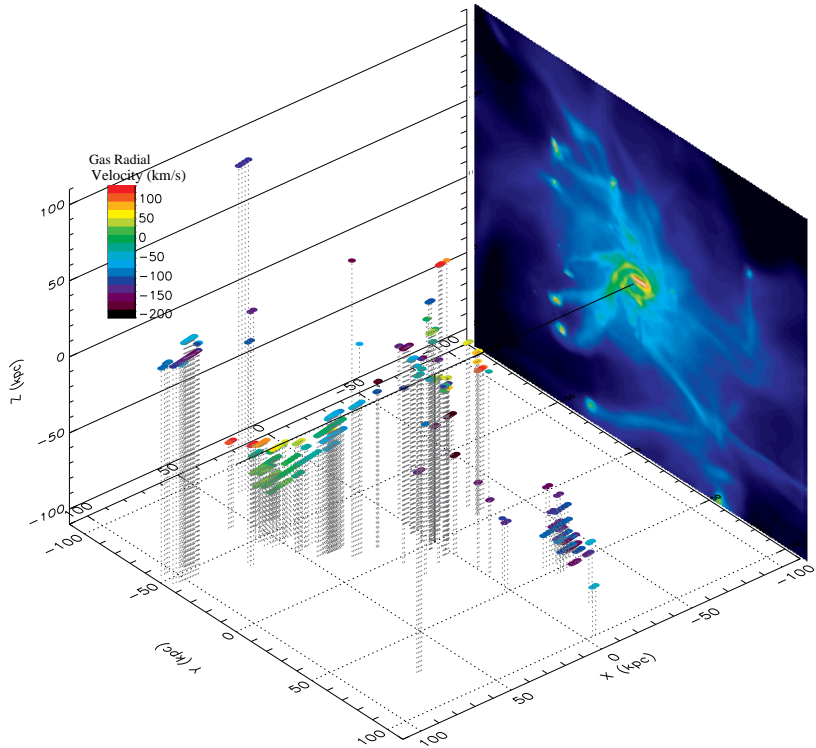}
\caption[angle=90]{Same as Figure~\ref{fig:plotVabs}, except the
  velocity color coding is for the radial velocity component with
  respect to the galaxy.  Red absorbing gas is outflow; blue is
  inflow. The majority of the {\MgII} absorption arises in filaments
  and tidal streams and is infalling towards the galaxy.}
\label{fig:plotVr}
\end{center}
\end{figure*}


\subsection{Disk Halo Models of The Simulated Galaxy}

Given the structures shown in Figures~\ref{fig:eonH}--\ref{fig:fonHI},
it would seem to be unrealistic to treat the halo as a monolithic
thick disk. However, we have modeled all the simulated {\MgII}
absorption profiles for each of the three galaxy inclinations with the
disk halo model (\S~\ref{halo}).

We find that for the edge--on view, out of 87 lines of sight with
detected {\MgII} absorption, 45\% have kinematics consistent with the
model (the full range of velocities can be explained). For
$i=45^{\circ}$, out of 124 lines of sight, 26\% have kinematics
consistent with the model. For the face--on view, out of 96 lines of
sight, only one is consistent with the model. This is not surprising,
since the projected maximum rotation velocity is small compared to the
absorption spreads.

The conclusion to be drawn here is, that even if the halo gas does not
rotate as a monolithic disk in the simulations, {\MgII} absorption
detected along some lines of sight can still appear to be consistent
with a disk halo model.

\begin{figure}
\begin{center}
\includegraphics[angle=0,scale=0.85]{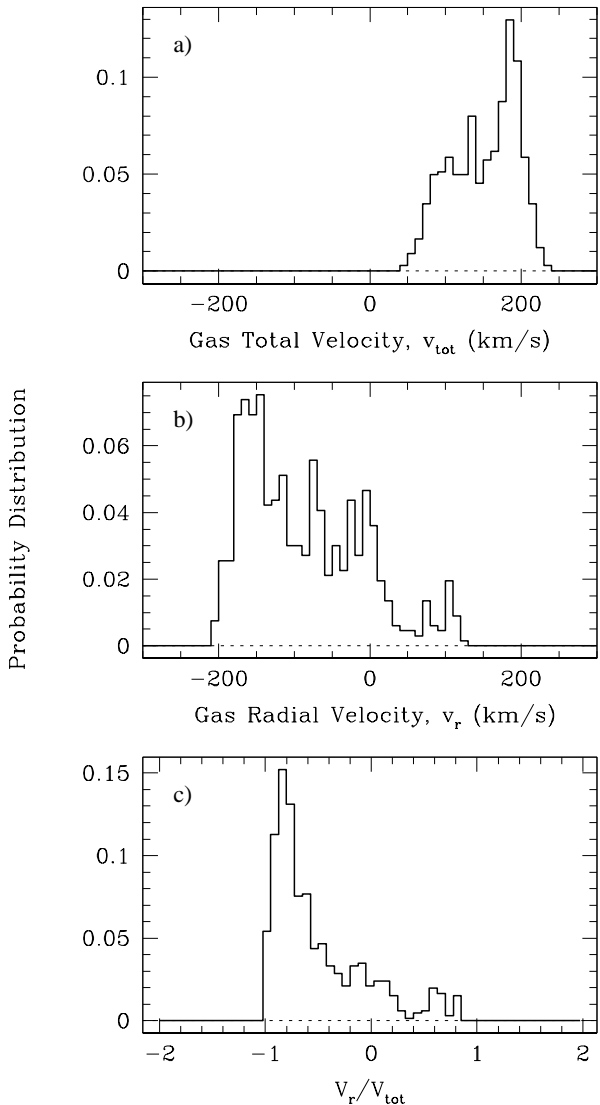}
\caption[angle=90]{Velocity distributions of {\MgII} absorbing gas cells within the 220~kpc cube of the simulation shown in Figures~\ref{fig:plotVabs} and \ref{fig:plotVr}. Gas within a $20$~kpc radius of the galaxy is omitted. --- (a) The total velocity distribution. --- (b) The radial velocity distribution. --- (c) The distribution of the ratio of radial to total velocity. A ratio of $v_r/v_{tot}=\pm 1$ indicates pure radial motion.}
\label{fig:plotVr_Vt}
\end{center}
\end{figure}

\subsection{Halo Gas Spatial and Velocity Distribution}
\label{sec:gas}

To understand what kinematic mechanisms are responsible for the
{\MgII} absorption velocity spreads measured in the simulated spectra,
we examine the velocities and spatial distributions {\MgII} absorbing
gas.

In Figure~\ref{fig:plotVabs}, we present the 3D spatial distribution of N(\MgII)$> 10^{11.5}$~{\cmsq} gas contributing to the {\MgII} absorption along the lines of sight. The edge--on galaxy is located at the origin and the lines of sight are parallel to the x--axis.  Gas with N(\MgII)$<10^{11.5}$~{\cmsq} is not shown for clarity (which corresponds to an equivalent width limit of 0.012~{\AA} for $b=5$~{\kms}).  From this view, the tidal streams and filaments can be visually discerned.  There are also isolated clouds that produce {\MgII} absorption.  Regions that contribute to absorption have physical sizes of $225$~pc (simulation resolution) to $\sim20$~kpc. All the absorption along the lines of sight occurs within $\pm110$~kpc from the center of the galaxy.

The gas is color coded as a function of line of sight velocity
relative to the galaxy systemic velocity. Gas that is colored red
(blue) is moving away from (towards) the observer.  The gas has line
of sight velocities ranging from $-160 \leq v_{los} \leq +160$~{\kms}.
The line of sight velocity dispersion in the inner regions near the
galaxy appears inconsistent with disk rotation. The somewhat
randomized velocities reflect the winds.  Upon carefully examining the
gas velocities along a particular line of sight, velocity gradients
can be observed. For example, the region seen in
Figure~\ref{fig:plotVabs} at $x=100,y=-50$~kpc, has velocity gradients
of $\sim30-40$~{\kms}.


In Figure~\ref{fig:plotVr}, we present the same three dimension
spatial distribution shown in Figure~\ref{fig:plotVabs}, except that
the gas is color coded as a function of radial velocity relative to
the galaxy. Gas colored red (blue) has a radial velocity component
that is outflowing from (infalling towards) the galaxy.  This combined
spatial and kinematic representation provides an holistic view of the
halo dynamics. The gas has radial velocities ranging from $-200 \leq
v_r \leq +100$~{\kms}.  If one focuses on the filament structure
($x=0$~kpc, $y=0-100$~kpc) in the plane of the galaxy, one can see the
strong velocity gradient. The gas along the filament far from the
galaxy increases in infall velocity from $\sim70$ to $\sim200$~{\kms}
as it approaches the galaxy center. The same dynamics can be seen
along the tidal stream ($x=0-50$~kpc, $y=-50$~kpc) originating from
the two small satellite galaxies.


In Figure~\ref{fig:plotVr_Vt}$a$, we show the probability distribution
of the {\it total\/} gas velocity of the {\MgII} absorbing gas cells.
What we call the probability distribution is the area normalized
frequency distribution that we detected in the simulations.  The total
velocity is the magnitude of the gas velocity vector relative to the
galaxy.  Since our observational data have only impact parameters
greater than $\sim20$~kpc, we exclude all absorbing gas within
$D\leq20$~kpc of the simulated galaxy. The velocities range from
$50\lesssim v_{tot}\lesssim 250$~{\kms} with a peak at $185$~{\kms}
and secondary maxima at $100$~{\kms}. In
Figure~\ref{fig:plotVr_Vt}$b$, we show the probability distribution of
radial velocities of the {\MgII} absorbing cells.  The velocities
range from $-200\lesssim v_{r}\lesssim +110$~{\kms} with a maximum at
$-160$~{\kms}. It appears that, beyond $D\geq20$~kpc, most of the gas
is infalling towards the galaxy and very little is outflowing. In
Figure~\ref{fig:plotVr_Vt}$c$, we show the ratio of the radial to the
total velocity. The bulk of the gas is dominated by radial infalling
velocities.

\begin{figure}
\begin{center}
\includegraphics[angle=0,scale=0.70]{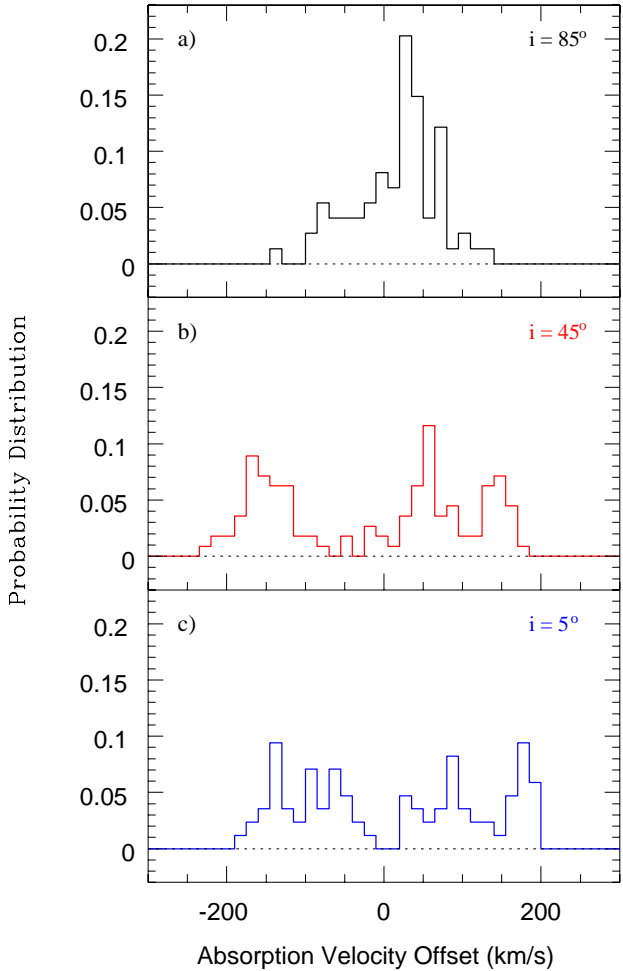}
\caption{Probability distributions of the mean velocities of the simulated {\MgII} absorption profiles arising in the lines of sight, for three galaxy orientations.  The quantity $v_{abs}$ is the optical depth weighted mean of the profiles.  Lines of sight with impact parameters less that 20~kpc are omitted. --- (a) $i=85^\circ$, edge--on.  --- (b) $i=45^\circ$. --- (c) $i=5^\circ$, face--on. The absorption profiles span the rotation velocities of the simulated galaxy ($v_c=180$~{\kms}).}
\label{fig:plotVoff}
\end{center}
\end{figure}

Drawing from the number of lines of sight we have through the
simulations, we produced the probability distribution of absorption
velocity offsets from the galaxy systemic velocity ($\Delta
v=v_{abs}-v_{gal}$~\kms) using the simulated absorption profiles. The
quantity $v_{abs}$ is obtained by calculating the optical depth
weighted mean of the profiles (the velocity at which there is equal
optical depth to both sides along the profiles).  In
Figures~\ref{fig:plotVoff}$a,b$, and $c$, we show the velocity offset
probability distributions for the edge--on, $i=45^\circ$, and face--on
orientations, respectively. For the edge--on case, the absorption
velocity offset ranges from $\pm100$~{\kms} with a strong peak around
$+30$~{\kms}; it is highly probable to detect absorption to one side
of the galaxy systemic velocity. For the $i=45^\circ$ case, the
velocity spread increases to about $\pm200$~{\kms} and develops
multiple peaks at $\pm150$~{\kms} and at $\sim+50$~{\kms}. The
probability of detecting gas at the galaxy systemic velocity is
significantly smaller than over the range $|50-150|$~{\kms}.  For the
face-on case, the velocity dispersion is still around
$\pm200$~{\kms}. Again, it is unlikely that absorption will be
detected at the galaxy systemic velocity.

Since most of the {\MgII} gas arises between $\pm200$~{\kms}, with
very little at the galaxy systemic velocity, and since most projected
galaxy rotation curves have maximum velocities of $\pm200$~{\kms}, it
may not be a surprise to observe {\MgII} absorption aligned with the
observed galaxy rotation curve.  These results are consistent with the
findings of \citet{bouche07}, who detected galaxy $\rm{H}\alpha$
emission within $\pm200$~{\kms} of the optical depth weighted mean
{\MgII} absorption redshift.

\subsection{Shortcoming of the Simulations}

The technique of quasar absorption lines through cosmological simulations is one of several promising approaches to understanding the dynamics of galaxy halos.  At the present time, simulations of galaxy formation in the cosmological context still need to achieve greater accuracy for modeling stellar feedback. For this study, we employed a feedback recipe that successfully results in extended metal enriched gas around galaxies.  These simulations result in an equivalent width distribution with an under abundance of larger equivalent widths and a relative over abundance of smaller equivalent widths.  They also under predict the observed {\MgII} mean absorption covering fraction range of $0.2-0.9$ \citep{tripp-china,chen08,kacprzak08,barton09}.  The covering fractions for all absorption above 0.02~{\AA} are as follows:  (1) $i=90^{\circ}$ (edge--on): total 10\%, weak 6\%, strong 3\%; (2) $i=0^{\circ}$ (face--on): total 10\%, weak 6\%, strong 3\% (3) $i=45^{\circ}$: total 14\%, weak 10\%, strong 5\%.  

These mismatches with observations could either be a result of the method in which {\MgII} column densities are determined in the simulations or observational biases.  In the simulations, the determination of the {\MgII} ionization fraction may be underestimated due to the fact that we do not account for shielding of UV photons in the ionization corrections.  It is also possible that the resolution of the simulations may influence the derived {\MgII} column densities and that higher resolution may in fact lead to higher column densities.  We aim to analyze such issues in future work.  A possible observational biases that galaxies are selected, identified, and assigned to already known absorption systems.  This may elevate the inferred covering fraction \citep[e.g.,][]{tripp-china}.
We do emphasize, however, that in the simulations we do detect {\MgII} absorption out to $\sim120$~kpc, as seen in current observations \citep{cwc-china,zibetti06,kacprzak08}.


As an additional caveat, we also remark that the experiment to examine the spatial and kinematic relationship between the galaxy and the {\MgII} absorbing structures in the simulations is very different than the observational experiment in one regard.  We examine a grid of sight lines through a single simulated galaxy, which is in a unique environment and undergoes a unique evolution in the IGM.  The observational data, on the other hand, sample a single line of sight through various galaxies in various environments and with various evolutionary histories and with random orientation through the galaxy and environment.  

We re--emphasize that the simulations, as we have applied them here, provide a first view of the types if physical structures and their spatial and kinematic relationship to a galaxy in the cosmological environment that give rise to {\MgII} absorption.  Though the details of the {\MgII} absorbing properties are not yet tuned to observations (and we mention that our experiment is only for a single galaxy using an incomplete sampling of its halo-- a full comparison with survey statistics is not entirely applicable), the dynamical structures themselves in the simulations are robust.  That we find some parts of these structures give rise to {\MgII} absorption is not outside expectation given the densities and temperatures of the gas.  The upshot is that, given these discrepancies and concerns, the simulations are not expected to provide a fully quantitative comparison with the data.  However, the simulations do provide a fully self--consistent galaxy model from which we acquired valuable insights for interpreting {\MgII} absorption line observations.

\subsection{Summary III: Galaxy Simulations and Halo Gas Distributions}

In our study of $z=0.923$ galaxy at three inclination angles with respect to the simulated quasar lines of sight, we have detected {\MgII} absorption in a variety of structures.  {\MgII} absorption was detected in inflowing metal enriched filaments, tidal streams, small satellites, and gas around the host galaxy. The absorption resides in a ``halo'' of about 100~kpc in size.  The types of structures that form in the simulations (see Figures~\ref{fig:eonH}--\ref{fig:fonHI}) are a challenge to models in which the absorption is assigned to thick disks \citep[e.g.,][]{cc96} and symmetrically distributed halos \citep[e.g.,][]{mo96}.  The structures also may provide further guidance for expanding upon radial density dependent halo occupation models \citep[e.g.][]{tinker08}.  The spatial distribution of filaments, tidal streams, and satellite galaxies are asymmetric, patchy (low volume filling factor), complex, and part of a cosmological setting.

DLA {\HI} column densities are seen out to $\sim100$~kpc, and arise
in low mass satellite galaxies. These galaxies are below the detection
limits of even deep {\it HST\/} images. Although, the covering
fractions of these dense regions are low, this might explain why some
bright galaxies at DLA redshifts are found at large impact parameters
\citep[e.g., the 3C 336 $z=0.656$ DLA at $D=100$~kpc;][]{s97}.
 
In the simulations, the kinematics are closely coupled to the gas
structures.  Metal enriched tidal streams and filaments are dominated
by infall to the central galaxy, and these structures are selected by
{\MgII} absorption (see Figures~\ref{fig:plotVabs} and
\ref{fig:plotVr}).  In fact, we find that gas giving rise to the
{\MgII} absorption is dominated by inflow (see
Figure~\ref{fig:plotVr_Vt}).  Inflow velocity gradients are
apparent, such that the infall increases as the gas approaches the
galaxy.  This inflow is not spatially symmetric. 

Similar to the observations, the velocity and spatial distributions
from the simulations conspire to give rise to {\MgII} absorption to
one side of the galaxy systemic velocity even though the absorbing gas
is not rotating with the star forming component (see
Figures~\ref{fig:stream} and \ref{fig:center}).  The velocity offset
probability distribution (relative to the simulated galaxy) spans
$\sim200$~{\kms} with lowest probability of detecting {\MgII} at the
galaxy systematic velocity (see Figure~\ref{fig:plotVoff}).  Thus, the
fact that we observe {\MgII} absorption velocities consistent with the
galaxy rotation curves may be a natural consequence of the spatial and
kinematic distributions of gas in complex environments surrounding
galaxies.

The current feedback recipes used in the simulations successfully
produces in extended metal enriched halo gas around galaxies. However,
for this single galaxy we derive covering fractions that are lower
than the current observed mean values of $0.2-0.8$
\citep{tripp-china,chen08,kacprzak08,barton09}. In this particular
galaxy halo, we also find an under abundance of larger equivalent
widths and a relative over abundance of smaller equivalent widths. We
have yet to determine if some of these mismatches are due to
observational biases in current studies or issues with the
feedback/baryon physics implemented in the simulations. However, the
structures in the simulations still provide valuable insight for
interpreting {\MgII} absorption line observations.


\section{Large Scale Galactic Winds: Observations and Simulations}

In our analysis of the simulated galaxy, we do not find systematic
rotation near the galaxy, as was suggested by \citet{lanzetta92}, but
that the gas kinematics reflects processes such as winds and chimneys.
These are characteristic of stellar feedback processed that
successfully circumvent the over cooling and angular momentum problems
that have plagued simulations \citep{ceverino09}.

Observationally, \citet{weiner08} has inferred outflows in {\MgII}
absorption associated with $z\sim 1$ star--forming galaxies ($10
\lesssim SFR \lesssim 40$~M$_\odot$~yr$^{-1}$). The outflow velocities
are proportional to $SFR^{0.3}$. The star formation rate in our
simulated galaxy is $3.5$~M$_\odot$~yr$^{-1}$. Calibrating to their
observations, our galaxy is expected to have winds in the range
$220-350$~{\kms}. Our simulated {\MgII} profiles indicated some
outflowing gas with velocities of $\sim 200$~{\kms} in the inner
$15$~kpc (independent of galaxy disk inclination). The absorption
profiles produced by the outflows are saturated and span both sides of
the systemic velocity of the simulated galaxy (see
Figure~\ref{fig:center}). Thus, the saturated {\MgII} absorption
profile that spans both sides of the galaxy systemic velocity might be
direct signature of outflows.

Two galaxies in our sample, Q$0002+051$ G1 and Q$0836+113$ G1, have
saturated profiles that span both sides of the galaxy systemic
velocity.  These saturated, symmetric profiles are also observed at
high redshift ($z\sim2$) and span the galaxy systemic velocity. The
absorption is attributed to large scale galactic outflows which can be
detected at least out to impact parameters of $\sim70$~kpc (Steidel
{\etal} 2010, in preparation). The outflows are determined to be more
or less symmetric and radial, giving rise to symmetric absorption
profiles. These high redshift galaxies do not exhibit substantial
kinematic structure and are instead consistent with
dispersion-dominated velocity fields around $80$~{\kms} \citep{law07},
similar to the shear observed for Q$0002+051$ G1 and Q$0836+113$
G1. Given the similarities between these two galaxies and those at
high redshift observed to have outflows, one possibility is that large
scale outflows are responsible for the absorption velocities
associated with both galaxies. To explore the outflow scenario for
Q$0002+051$ G1 and Q$0836+113$ G1, we examine the galaxy star
formation rates.

For Q$0836+113$ G1, we measure a {\OII} line flux of $3.4\pm0.43\times
10^{-16}$~erg~s$^{-1}$~cm$^{-2}$. We use the relation from
\citet{kewley04} to obtain a star formation rate of
6.5~M$_{\odot}$yr$^{-1}$. For Q$0002+051$ G1 we determined the star
formation from the UV flux at 1700~{\AA} which was derived from the
$R$--band magnitude of the galaxy \citep{sdp94}. The robustness of this
method has been demonstrated by \citet{erb06}. From the measured UV
flux of $2.3\pm0.5\times 10^{-16}$~erg~s$^{-1}$~cm$^{-2}$, we derive a
SFR of 6.5~M$_{\odot}$yr$^{-1}$. The SFRs are not corrected for the
internal extinction of the galaxies and are thus lower limits.

Using the results of \citet{weiner08}, both galaxies are expected to
have outflows with velocities in the rage $270-430$~{\kms}. Both
galaxies are at impact parameters of $\sim25$~kpc, which is near the
boundary of where we lose the signature of outflows for our simulated
galaxy.  Although Q$0836+113$ G1 and Q$0002+051$ G1 have {\MgII}
absorption velocity widths of 290 and 470~{\kms}, respectively, which
are comparable to the outflow velocity range predicted from the
results of \citet{weiner08}.

\citet{heckman02,heckman03} discusses that the star formation per unit 
area is a more relevant indicator of galaxy outflows. These outflows
are ubiquitous in galaxies where the global star-formation rate per
unit area exceeds $\Sigma=0.1$~M$_{\odot}$~yr$^{-1}$~kpc$^{-2}$, where
the area is defined by the half light radius of the galaxy. This
criteria applies to local starbursts and even high redshift Lyman
Break galaxies. The ISM entrained in the winds have outflow speeds of
$\sim100$ to $\sim1000$~{\kms}.  For Q$0002+051$ G1 we obtain a
$\Sigma \geq 0.35$~M$_{\odot}$~yr$^{-1}$~kpc$^{-2}$. For Q$0836+113$ G1 we
obtain a $\Sigma \geq 0.14$~M$_{\odot}$~yr$^{-1}$~kpc$^{-2}$.  Thus, both
of the galaxies are expected to have outflow signatures.

We estimate that the simulated galaxy has a
$\Sigma=0.08$~M$_{\odot}$~yr$^{-1}$~kpc$^{-2}$, which is slightly less
that the criterion of Heckman.  This is consistent with our outflow
velocities derived from the {\MgII} absorption profiles since we do
not see evidence of strong large scale outflows. This particular
simulated galaxy may not be well representative of the Q$0002+051$ G1
and Q$0836+113$ G1 galaxies.

In summary, galaxies Q$0002+051$ G1 and Q$0836+113$ G1 are
kinematically similar to high redshift absorption selected galaxies. The
SFRs and $\Sigma$s for both galaxies exceed the limits were strong
outflows are expected. Given the large impact parameters that outflows
are detected at high redshift, it is quite possible the observed
{\MgII} absorption kinematics for galaxies Q$0002+051$ G1 and
Q$0836+113$ G1 are signatures of outflowing gas.


\section{Conclusions}
\label{sec:conclusionkine}

We have examined and compared the detailed galaxy and {\MgII}
absorbing kinematics for a sample of 10 intermediate redshift
galaxies. The galaxies have a wide range of inclinations and
orientations with respect to the background quasar. The galaxy--quasar
impact parameters range from $26 \leq D \leq 108$~kpc.  The galaxy
rotation curves were obtained from ESI/Keck spectra and the {\MgII}
absorption profiles were obtained from HIRES/Keck and UVES/VLT quasar
spectra. In an effort to compare the relative kinematics, we used a
thick disk halo model to compute the expected absorption velocities
through a monolithic gaseous halo.

To obtain theoretical insights into the gas dynamics and spatial
distribution of halos, we used the technique of quasar absorption
lines to analyze {\MgII} absorption around a galaxy in a high
resolution cosmological simulation of galaxy formation. The galaxy was
probed with a square grid of sightlines at intervals of 7.5 kpc that
span $-110$ kpc to $+110$ kpc for a total of 900 sightlines.  We
examined this galaxy at three different inclinations, face--on,
$45^{\circ}$, and edge--on.
 
Our mains results can be summarized as follows:

\begin{enumerate}

\item
For all ten galaxies, the velocity of the strongest {\MgII} absorption
component lies in the range of the observed galaxy rotation curve.  In
seven of ten cases, the {\MgII} and {\MgI} absorption velocities
reside fully to one side of the galaxy systemic velocity. The
strongest absorption usually aligns with one arm of the rotation
curve. In the three remaining cases, the absorption velocities span
both sides of the galaxy systemic velocity. Two of those three
(Q$0002+051$ G1 and Q$0836+113$ G1) have strong saturated absorption
on both sides of the galaxy systemic velocity.  The third (Q$1127-145$
G3), has two very weak clouds, and therefore probes low column density
gas.

\item
For galaxies Q$0002+051$ G1 and Q$0836+113$ G1, we have determined
that large scale galactic outflows might be giving rise to the
observed {\MgII} absorption kinematics. Both galaxies have a $SFR \sim
6.5$~M$_\odot$~yr$^{-1}$ and $\Sigma \geq
0.35$~M$_{\odot}$~yr$^{-1}$~kpc$^{-2}$ and $\Sigma \geq
0.14$~M$_{\odot}$~yr$^{-1}$~kpc$^{-2}$, respectively. These SFRs and
$\Sigma$s are typically found for galaxies exhibiting outflow
velocities of several hundred {\kms}.  The {\MgII} absorption
velocities associated with the two galaxies span both sides of their
systemic velocity. Such profiles have been interpreted, in both our
simulations and at high redshift, as signatures of outflows.

\item
We find that the observed {\MgII} absorption velocity spread and
optical depth distribution may be a function of galaxy inclination.
Galaxies with higher inclination exhibit a {\MgII} absorption velocity
spread of $\sim300$~{\kms} with a somewhat even distribution of
optical depths, whereas, galaxies with lower inclinations exhibit a
narrower velocity spread of $\sim100$~{\kms} with a clear optical
depth peak at $v\sim100$~{\kms}. A K-S test shows that the {\MgII}
optical depth distributions for the high and low inclination bins are
not consistent at the $3.3\sigma$ level. These results suggest that
the absorbing gas is either disk--like or the spatial distribution and
kinematics of the structures producing the absorption (i.e.,
filaments, tidal streams, satellites etc.) are closely coupled to the
disk orientation.

\item We employed simple rotating disk halo models to examine whether disk--like rotation is consistent with the observed galaxy--gas kinematics.  For model parameters that allow for a 1 Mpc gas scale height and maximum rotation velocity (rigid rotation) the the bulk of the observed absorption kinematics can be explained by co--rotation with the galaxy.  In all cases, the rotating disk halo models we present are unable reproduce the {\it full} spread of observed {\MgII} absorption velocities.  This model is a highly unrealistic representation of galaxy gas.  When the parameters are relaxed to better reflect reasonable gas scale heights and a slowing of the rotation speed with height above the disk plane, the relative proportion of the gas velocity spread that can be made consistent with galaxy co--rotation diminishes such that some absorbers cannot have but a tiny fraction explained by co--rotation.  In this simple scenario, even if some of the absorbing gas arises in a thick disk, what we learn from the exercise is that some additional type of dynamical process (such as infall, outflow, supernovae winds, etc.)  must be invoked to explain the range of absorption velocities hat cannot be made consistent with the simple rotating disk halo model.

\item 
In two quasar fields, we find pairs of galaxies that align in velocity
within $~\sim100$~{\kms} of a single, saturated {\MgII} absorption
system.  For one case, the observed velocity range of the strong
saturated component can be explained by a rotating disk model only if
{\it both\/} galaxies contribute to the absorption.  This challenges
the idea that an individual {\MgII} absorber can be assigned to a
single galaxy, and understood as an isolated halo.

\item
In the simulations, {\MgII} absorption selects gas structures such as
metal enriched tidal streams, filaments, small satellite galaxies, and
the region within $\sim20$~kpc of the galaxy. Together, these
structures extend roughly $\sim100$~kpc around the galaxy, suggesting
that galaxy ``halos'' are a complex composite of the these various
structures.

\item For this simulated galaxy the {\MgII} covering fraction is $\sim
10$\%, which is below the current observational estimated means of
20--80\% \citep{tripp-china,kacprzak08,barton09}.  This may reflect a
need for additional tuning of the feedback/baryon physics in the
simulations, or indicates current observational biases.

\item
In the simulations, DLA {\HI} column densities arise in low mass
satellite galaxies at impact parameters as large as
$\sim100$~kpc. These galaxies are below the detection limits of deep
{\it HST\/} images. Although, the covering fractions of these dense
regions are low, this might explain why some bright galaxies at DLA
redshifts are found at large impact parameters.

\item
In the simulations, the majority of the {\MgII} absorbing gas is
infalling in filaments and tidal streams towards the galaxy with
velocities between $-200 \leq v_r \leq -180$~{\kms}.  The velocity
offset probability distribution (relative to the simulated galaxy)
spans $\sim200$~{\kms} with lowest probability of detecting {\MgII} at
the galaxy systematic velocity. Thus, observed {\MgII} absorption
velocities can fall within the range of the galaxy rotation curve
velocities, even though the gas arises in a variety of kinematics
structures.

\end{enumerate}

The gas structures selected by {\MgII} in the simulations (see Figures~\ref{fig:eonH}--\ref{fig:fonHI}) cannot be described as simple thick disks or spherical halos.  If the simulations reflect reality, it would appear that {\MgII} absorption arises in large $\sim 100$~kpc halos that are built from the local cosmological environment of a moderate mass galaxy.   Complicating the picture is the fact that we find groups and pairs of galaxies that align in velocity within $~\sim100$~{\kms} of a single {\MgII} absorption system. This challenges the idea that an individual {\MgII} absorber can be assigned to a single galaxy or understood as an isolated halo. Though considered subcomponents of halos, smaller scale structures like the Magellanic-type galaxies and tidal streams, may contribute significantly to the detections of {\MgII} absorption \citep{york86,kacprzak07}. These considerations lead us to suggest that galaxies and {\MgII} absorbers should be studied and modeled in a environmental context if they are to be fully understood. 

In the simulations, the kinematics are closely coupled to the gas
structures (i.e., filaments, tidal streams, small satellite galaxies,
and the inner 20~kpc of the central galaxy).  As observed in our data,
the simulated {\MgII} absorption velocities fall within the range of
the galaxy rotation velocities, and rarely at the galaxy systematic
velocity.  Thus, the simulations suggest that observing {\MgII}
absorption velocities consistent with the galaxy rotation curves can
naturally occur even if the absorption arises in many different
structures in the complex environment of the galaxy.  It is these
structures that comprise halos.

A natural extension of the work presented here would be to perform a
similar study (simulations and observations) that incorporates the
kinematics of higher ionization {\CIVdblt} and {\OVIdblt} doublet
absorption.  These ions probe lower density and/or higher temperature
structures and provide a more comprehensive view of the gaseous
environment around galaxies.  Future observations with the Cosmic
Origins Spectrograph are perfectly suited for the galaxy sample
presented in this paper.  It is also important to expand the number of
galaxy environments studied in the simulations.

\acknowledgments 

We thank Greg Wirth for his help and advice with ESI/Keck.  We are
grateful to A. Kravtsov for providing the hydro code. We are in debt
to N. Gnedin creating the graphics package IFRIT. We thank Aneta
Siemiginowska for her discussion regarding the X--ray data of
Q$127-145$. We express our gratitude to the anonymous referee for a
careful reading and for insightful comments that lead to an improved
manuscript. C.W.C and G.G.K were funded by the NSF grant AST
0708210. G.G.K was partially funded by the NMSU Graduate Research
Enhancement Grant.  M.T.M thanks the Australian Research Council for a
QEII Research Fellowship (DP0877998).  Most of the data presented
herein were obtained at the W.M. Keck Observatory, which is operated
as a scientific partnership among the California Institute of
Technology, the University of California and the National Aeronautics
and Space Administration. The Observatory was made possible by the
generous financial support of the W.M. Keck Foundation.  Some
observations were made with the NASA/ESA Hubble Space Telescope,
obtained from the Data Archive at the Space Telescope Science
Institute, which is operated by the Association of Universities for
Research in Astronomy, Inc., under NASA contract NAS 5--26555.  Some
of this research was based on observations made with ESO Telescopes at
the Paranal Observatories under program IDs listed in
Table~\ref{tab:qsospec}. The computer simulations presented in this
paper were performed at the National Energy Research Scientific
Computing Center (NERSC) of the Lawrence Berkeley National Laboratory.





{\it Facilities:} \facility{HST (WFPC--2)}, \facility{Keck II (ESI)},
\facility{Keck I (HIRES)}, \facility{VLT (UVES)}.

\end{document}